\documentclass[apj]{emulateapj-rtx4}
\usepackage{amsmath}
\usepackage{amssymb}
\usepackage{apjfonts}

\newcommand{\cmc}{\,{\rm cm$^{-3}$}\,}
\newcommand{\kms}{\,{\rm km\,s$^{-1}$}\,} 
\newcommand{\kmsmpc}{\,{\rm km\,s$^{-1}$\,Mpc$^{-1}$}\,}

\newcommand{\etal}{{ et~al.~}}

\newcommand{\photflux}{\,{\rm photons\,s$^{-1}$\,cm$^{-2}$}\,}

\newcommand{\ergs}{\,{\rm erg\,s$^{-1}$}\,}
\newcommand{\ergscm}{\,{\rm erg\,s$^{-1}$\,cm$^{-2}$}\,}
\newcommand{\Ms}{M_\odot}
\newcommand{\Msyr}{\,M_\odot\,{\rm yr}^{-1}\,}
\newcommand{\Zs}{Z_\odot}
\newcommand{\Ls}{L_\odot}
\newcommand{\chidof}{\chi^2/{\rm dof}\,}
\newcommand{\ergscmc}{\,{\rm erg\,cm$^{-3}$}\,}
\newcommand{\cnts}{\,{\rm counts\,s$^{-1}$}\,}
\shorttitle{Edges and Bubbles in NGC\,5846}
\begin{document}

% ------------------ title page ---------------------------------------------

\title{Deep {\it Chandra} Observations of Edges and Bubbles in the
  NGC\,5846 Galaxy Group} 

\author{ 
Marie E. Machacek$^1$, Diab Jerius$^1$, Ralph Kraft$^1$, 
William R. Forman$^1$, Christine
  Jones$^1$, Scott Randall$^1$, Simona Giacintucci$^2$, Ming Sun$^3$ 
}
\affil{$^1$ Harvard-Smithsonian Center for Astrophysics \\ 
       60 Garden Street, Cambridge, MA 02138 USA}
\affil{$^2$ Department of Astronomy, University of Maryland, 
      College Park, MD 20742-2421 USA}
\affil{$^3$ Department of Astronomy, University of Virginia, 
     P.O. Box 400325, Charlottesville, VA 22901, USA}
\email{mmachacek@cfa.harvard.edu}

%--------------------------------------------------------------------
\begin{abstract}
We use a combined $120$\,ks {\it Chandra} exposure to analyze 
X-ray edges produced by non-hydrostatic gas motions (sloshing) 
from galaxy collisions, and cavities formed by AGN
activity. Evidence for gas sloshing is seen in the
spiral morphology and multiple cold front edges in NGC\,5846's 
X-ray surface brightness distribution, while lack of   
spiral structure in the temperature map suggests the perturbing 
interaction was not in the plane of the sky.  Density and spectral 
modeling across the edges indicate that the relative motion of gas in 
the cold fronts is at most transonic. Evidence for AGN activity is 
seen in two inner bubbles at $0.6$\,kpc, filled with 
 $5$\,GHz and $1.5$\,GHz radio plasma and coincident with H$\alpha$
emission, and in a ghost bubble at $5.2$\,kpc west of NGC\,5846's
nucleus. The outburst energy and ages for the inner (ghost) 
bubbles are $\sim 10^{55}$\,ergs and $\sim 2$\,Myr 
($\sim 5 \times 10^{55}$\,ergs and $12$\,Myr),respectively, implying
an AGN duty cycle of $10$\,Myr. The inner bubble rims are
threaded with $9$ knots, whose total $0.5-2$\,keV X-ray 
luminosity is $0.3 \times 10^{40}$\ergs, a factor $\sim 2-3$ less than 
that of the surrounding rims, and $0.7$\,keV mean temperature is 
indistinguishable from that of the rims. We suggest that the knots may
be transient clouds heated by the recent passage of a shock from the 
last AGN outburst. We also observe gas stripping 
from a cE galaxy, NGC\,5846A, in a $0.5$\,kpc long ($\sim
10^5\Ms$) hot gas tail, as it falls towards NGC\,5846.
\end{abstract}


% ----------------- start of body ------------------------------------------

\section{INTRODUCTION}
\label{sec:introduction}

Over the past decade, deep X-ray observations of significant
substructure in the hot gas near the centers of rich clusters 
have revolutionized our understanding of the evolution of clusters 
and the galaxies at their cores. Sharp surface
brightness discontinuities (edges) are found where the increase in 
density across the discontinuity is accompanied by a decrease in 
temperature, opposite to that expected from shocks. These `cold
fronts'  provide evidence
for an ongoing subcluster merger or non-hydrostatic motions of the 
gas (`sloshing') from a prior 
disturbance in the core (see, e.g. the review by Markevitch 
\& Vikhlinin 2007). Furthermore, episodic outbursts from the active 
galactic nucleus (AGN) at the centers of the dominant
galaxies in groups and clusters, are revealed by outflow shocks and  
X-ray cavities (deficits in X-ray surface brightness) in the 
 intra-cluster medium (ICM) (see, e.g. Fabian \etal 2003 for 
the core of the Perseus cluster, Forman \etal 2005 for M87 
in the Virgo Cluster). These X-ray cavities  correspond to regions where 
the hot cluster gas has been evacuated by jets from AGN outbursts 
and the interior of the cavity (bubble) is filled with radio 
plasma. After an initial momentum driven formation stage, the
'bubble' of radio plasma rises buoyantly in the group/cluster
atmosphere. This feedback between the central AGN and its environment 
may be key to resolving the 
cooling flow problem in galaxy clusters, and and may be an  
essential ingredient in the 
emerging models for the coevolution of galaxies and their central 
supermassive black holes (e.g. Best \etal 2006; Croton \etal 2006). 
While these phenomena have been 
studied in high angular resolution, deep X-ray observations for  
many rich clusters (e.g. McNamara \etal 2000; 
Nulsen  \etal 2002; Birzan \etal
2004; Nulsen \etal 2005; McNamara \etal 2005; Allen \etal 2006), 
their impact on structure formation in the much more common cores of galaxy 
groups has received far less attention (however, see e.g. 
Machacek \etal 2010 for the Telescopium group; Gitti \etal 2010 for
HCG\,62; Randall \etal 2011 for the NGC\,5813
group; David \etal 2011 for the NGC5044 group).  
Yet, it is in groups at high redshift where galaxy evolution is most 
rapid (Cooper \etal 2006), gas-rich mergers are more common, and,
since the gravitational
potential is shallower, outbursts from supermassive black holes
(SMBHs) may have a greater effect on the surrounding gas than in 
their richer cluster cousins.

\begin{figure}[t]
\begin{center}
\includegraphics[width=3.0in]{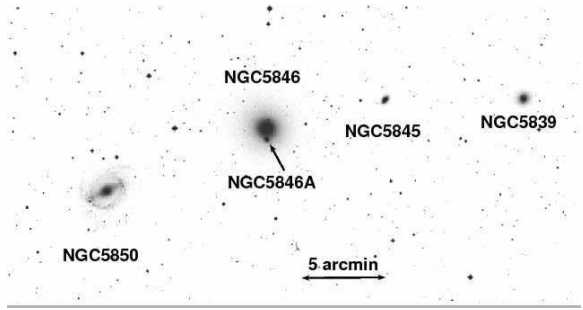}
\caption{{\footnotesize DSS image Bj band image of the NGC\,5846 group 
taken with the UK Schmidt Telescope on 1980 March 20. North is up and
East is to the left.}
}
\label{fig:n5846dss}
\end{center}
\end{figure}

In this paper we present results from the combined $\sim 120$\,ks  
{\it Chandra} observation of the NGC\,5846 group. The Digital Sky
Survey (DSS) image is shown in Figure 
\ref{fig:n5846dss}. Since the NGC\,5846
group is only $\sim 24.9^{+2.4}_{-2.2}$\,Mpc distant 
(Tonry \etal 2001), this exposure is one of the 
deepest high angular resolution  X-ray observations of any nearby 
group of galaxies. The NGC\,5846 group consists of $\sim 20$
bright (${\rm M}_{\rm B} \lesssim -14.9$) galaxy members with a group velocity 
dispersion of $386^{+51}_{-46}$\kms (Zabludoff \& Mulchaey 1998; 
Mulchaey \etal 2003).  Previous X-ray studies using the  Einstein
X-ray Observatory 
(Forman, Jones \& Tucker 1985; Biermann, Kronberg \& Schmutzler 1989) 
and using  ROSAT and ASCA (Finoguenov et al. 1999 and Trinchieri et
al. 1997), show $\sim 1$\,keV diffuse group X-ray emission extending 
to $\sim 200$\,kpc ($\sim 28\farcm7$) from the central, brightest group 
galaxy NGC\,5846 and a total X-ray luminosity within $200$\,kpc of 
$\sim 10^{42}$\ergs (Mulchaey \etal 2003). From combined optical and
X-ray studies using ROSAT, ASCA and {\it XMM-Newton} data, 
Mahdavi, Trentham \& Tulley (2005) found  that the 
NGC\,5846 group may be  part of a larger ($8 \times 10^{13}\Ms$) 
structure, sharing a common $1.6$\,Mpc diameter dark matter halo with the
NGC\,5813 group located $600$\,kpc away in projection, and that 
group membership is overwhelmingly dominated by dwarf elliptical 
galaxies ($80\%$), similar to that seen in the Virgo cluster. 
There is, however,  no evidence for interaction between the NGC\,5846
 and NGC\,5813 groups (Randall \etal 2011). 

 NGC\,5846 is a giant E0 elliptical galaxy with an effective radius 
of $11.6$\,kpc (Bender \etal 1992) and an absolute V magnitude of $-22.6$ 
(Faber et al. 1989), comparable to M87 (M$_V=-22.5$), the giant
elliptical galaxy in the Virgo Cluster. However, in contrast to M87
whose central velocity dispersion suggests a  black hole mass of at
least $\sim 3 \times 10^9\,\Ms$, NGC\,5846's central velocity dispersion
of $278$\,\kms corresponds to a supermassive black hole mass of
$3.5 \times 10^{8}\,\Ms$, an order of magnitude smaller 
(Tremaine et al. 2002; Gebhardt \& Thomas 2009). 
Radio emission from the nuclear region of NGC\,5846 is complex, with  
a central compact, possibly variable, steep spectrum ($\alpha = 0.8$) 
radio source with radio fluxes of $1.5$\,mJy ($2.8$\,mJy) at 
$5$\,GHz ($2.3$\,GHz), likely associated with an AGN. Three additional 
radio components are seen  in $2.3$, $5$, and $15$\,GHz VLBA maps
roughly aligned in the north-south direction 
with the nuclear emission (Filho \etal 2004). These regions, extended on 
milli-arcsecond scales,  may either be compact supernova remnants or emission 
from a jet (Filho \etal 2004). 
{\it XMM-Newton} observations
suggest at least $3$  surface brightness discontinuities (edges) 
in the gas surrounding NGC\,5846 (Finoguenov \etal 2006). Such
discontinuities indicate gas motions. These may be residual gas
`sloshing' from a previous galaxy merger or from a  
high velocity galaxy encounter with the large, 
HI rich, barred spiral galaxy NGC\,5850, located $10\farcm3$ 
($71.5$\,kpc in projection) to the east (see Fig. \ref{fig:n5846dss}). 
The large shift of HI gas to the west and northwest in NGC\,5850, in the
direction of NGC\,5846,
and NGC\,5850's distorted stellar arms and spurs, also to the west,
suggest a recent ($\lesssim 200$\,Myr) tidal interaction between the
two galaxies (Higdon \etal 1998).

Non-hydrostatic gas motions also could be induced by AGN activity. 
A  previous short ($30$\,ks)  {\it Chandra} X-ray observation of 
NGC\,5846 shows bright, knotty-rimmed X-ray cavities associated with 
radio lobes in NGC\,5846's  central $\sim 2$\,kpc (Trinchieri \& 
Goudfrooij 2002; Allen \etal 2006). Extensive filamentary 
H$\alpha$ + [NII] line emission and dust also are found in the central 
region of the galaxy, highly correlated with the X-ray emission 
(Goudfrooij \& Trinchieri 1998). Spitzer MIPS observations of
NGC\,5846 show extended $70\,\micron$ dust emission to a 
FW$0.1$M~$\sim 5$\,kpc, suggesting that dust, originating from 
stellar mass loss from evolved AGB stars, accumulated in the cental 
$\sim 1$\,kpc of that galaxy and was subsequently transported
buoyantly to these larger radii by AGN activity (Temi \etal 2007).   

In this paper we use the combined $120$\,ks Chandra exposure to study
in detail the X-ray cavities and surface brightness edges that
chronicle AGN activity in NGC\,5846 and the relative motions between  
the galaxy ISM and group gas. In \S\ref{sec:obs} we summarize 
our observational data reduction and analysis techniques. 
In \S\ref{sec:outerregions} we analyze the large scale surface brightness and 
temperature structure in the core of the NGC\,5846 group, 
discuss evidence for gas sloshing, and model the density, temperature 
and pressures across observed surface brightness edges. 
In \S\ref{sec:bubbles} we focus on the properties of the system of AGN 
outburst cavities near the center of NGC\,5846.  In \S\ref{sec:N5846A}
we also present the first observation of gas stripping from a compact
elliptical galaxy, NGC\,5846A, during its' infall towards NGC\,5846. 
Our conclusions are summarized in \S\ref{sec:conclude}. 
Unless otherwise indicated,
quoted uncertainties are $90\%$ CL for spectra and $1\sigma$ for
photon counts. For the standard $\Lambda$
dominated cold dark matter cosmology ($H_0 = 71$\kmsmpc, $\Omega_m =
0.27$, $\Omega_\Lambda = 0.73$) and assuming NGC\,5846 is at rest
at the center of the group potential, the luminosity distance to NGC\,5846
is $24.2$\,Mpc, and $1'' = 0.116$\,kpc (Wright 2006). This is
consistent within uncertainties with the luminosity distance
determined from surface brightness fluctuations (Tonry 2001). 

%--------------------------------------------------------------------------

\section{OBSERVATIONS AND DATA REDUCTION}
\label{sec:obs}

NGC\,5846 was observed with the {\it Chandra} X-ray Observatory 
for $30313$\,s on 2000 January 24 (obsid 788) in FAINT mode with
ACIS-S at aimpoint, and for $90367$\,s on 2007 June 12 (obsid 7923) 
in VFAINT mode with ACIS-I at aimpoint for a total of $\sim 120$\,ks. 
The data were initially filtered to reject bad pixel grades
($1$, $5$, $7$) and data that fell on hot pixels. Data flagged in 
obsid 7923 (VFAINT mode) as having excessive counts in the border
pixels surrounding event islands were also removed to optimize signal
to noise at energies below $1$\,keV. We then reprocessed these data 
with CIAO4.2 using standard  tools and calibrations provided by the 
Chanda X-ray Center, applying corrections for the charge transfer
inefficiency on the ACIS CCDs, the time-dependent build-up of
contaminant on the optical filter, and the secular drift 
(tgain\footnote{see,
  http://cxc.harvard.edu/contrib/alexey/tgain/tgain.html}) 
of the average pulse-height amplitude for photons of fixed
energy. Periods of anomalously high particle count rates (flares) were 
excised  using lc\_clean and a $3\sigma$ clipping algorithm, 
resulting in useful exposures of $23205$\,s and  $89509$\,s  for 
obsid 778 and obsid 7923, respectively. 

Backgrounds for the imaging and spectral analyses, where indicated,   
were drawn from the source free background data set\footnote{see
  http://cxc.harvard.edu/contrib/maxim/acisbg}, i.e. period C ACIS-S (
period D ACIS-I) for OBSID 788 (7923), appropriate for our
observation and instrument configuration, using the CIAO 
acis\_bkgrnd\_lookup tool. These data were reprocessed using the same
gain files as used for the source data, reprojected to the aimpoint of the
observation, and then normalized by comparing count rates in 
the $9.0-12.0$\,keV energy band, where particle background dominates. 
This resulted in additional normalization factors of $0.94$ and $1.44$ 
for obsid 788 and obsid 7923, respectively. Since NGC\,5846 lies close
to the Northern Spur, there is an additional soft Galactic background 
component not accounted for in the blank sky background data. This
Galactic component is expected to be uniform over the combined ACIS 
fields-of-view (FOVs). We  model this background using the 
flux measured in a $63\farcs55$ radius  circular region on CCD6
centered at (${\rm RA,\,Dec}\,=\,15^h06^m56.391^s$,
  $+01^\circ20'33\farcs66^s$), a distance of $17\farcm16$ from 
NGC\,5846 where group emission is expected to be small. 
This soft Galactic background component was also
subtracted from the images analysed in \S\ref{sec:outerregions}.
Instrument and exposure maps were generated in $0.25$\,keV passband 
intervals from $0.5$ to $4.0$\,keV and $2$\,keV intervals from 
$4.0 - 10.0$\,keV for both data sets using standard CIAO
tools. Background-subtracted, exposure corrected images were
constructed in various bandpasses and the data sets coadded for
further analysis. 

The CIAO tool wavdetect was run on the coadded, background-subtracted 
$0.5-4.0$\,keV image to identify point sources, 
and the source regions were adjusted to eliminate overlapping
regions.  An ellipse slightly larger than the source region was 
generated as a background region for each source. Sources outside the 
central $2.3$\,kpc ($20''$) of NGC\,5846, i.e. those not associated 
with knots on the inner bubble rims or NGC\,5846's nucleus, were
excised from the data. The point source regions in the 
exposure-corrected images were then filled with events sampled from
the corresponding background region around each source to produce the 
images of diffuse gas shown in \S\ref{sec:outerregions}.

Spectra were extracted from the point source cleaned event files 
using standard CIAO tools and fit with XSpec 11.3.0. 
Local background regions drawn from the ACIS-I detector are used, when
possible, to mitigate the effects of the soft Galactic background. 
The spectra from both observations were fit jointly, when such a fit
reduced uncertainties in the model parameters. Otherwise, we present
only the long $89$\,ks exposure (OBSID 7923) for our spectral
analysis, because of its higher number of source counts.

%-------------------------------------------------------------------------
\section{Large Scale Structures}
\label{sec:outerregions}

\begin{figure}[t]
\begin{center}
\includegraphics[width=3in]{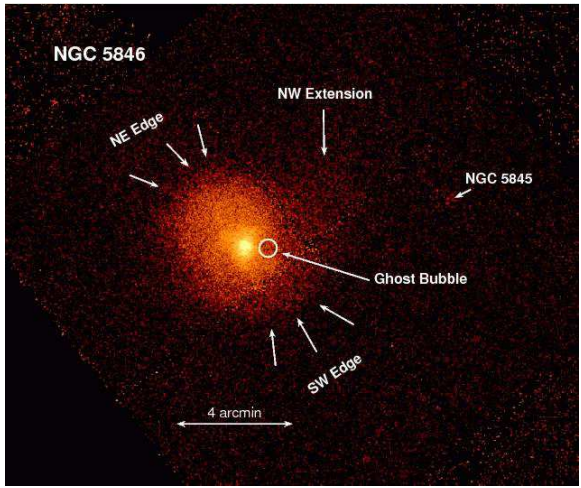}
\caption{{\footnotesize {\it Chandra} background-subtracted, 
 exposure-corrected, coadded  $0.75-2.0$\,keV image of
 diffuse gas in the outer regions of NGC\,5846 showing the 
 main X-ray features outside $2$\,kpc of the nucleus. 
$1 {\rm pixel} = 1\farcs0 \times 1\farcs0$ and the image has been
 smoothed with a $2\farcs0$ gaussian kernel to highlight faint features.
North is up and east is to the left.
}
} 
\label{fig:n5846full}
\end{center}
\end{figure}

In Figure \ref{fig:n5846full} we show the $0.75-2$\,keV X-ray
background subtracted, exposure corrected, coadded image
of diffuse gas in and surrounding NGC\,5846. The image has been
binned by $2$ pixels into $\sim 1''$ bins, 
and smoothed with a $2\farcs0$ Gaussian kernel to
highlight faint emission features outside the nuclear region of
NGC\,5846. The surface brightness distribution is brighter (gas
denser) from $\sim 4 - 20$\,kpc northeast of the nucleus than at
similar distances to the southwest. Surface brightness
discontinuities (edges) are found $\sim 2\farcm7$ ($\sim 19$\,kpc) 
to the northeast and southwest of NGC\,5846's nucleus, with 
possible inner edges to the southwest at $\sim 1\farcm5$ 
($\sim 10.5$\,kpc) and to the northeast at $\sim 1'$ ($\sim
6.7$\,kpc). In \S\ref{sec:densjump} we find that the ratio of 
 gas density across these  edges exceeds $1$ at $> 90\%$ confidence 
 level in all cases, confirming that these discontinuities are significant.
 We also observe a tail-like feature of enhanced emission (NW
 extension) at $13\sigma$ significance in the flux over expected 
 background, that extends $4\farcm1$ ($28.5$\,kpc) to the northwest  
 suggesting a spiral pattern. A roughly spherical deficit in X-ray
 surface brightness is found, at an $8\sigma$ significance in 
 flux compared to neighboring regions, located  $\sim 45''$ ($5.2$\,kpc) west
 of the nucleus. In contrast to the inner X-ray cavities observed in
 previous studies (Trinchieri \& Goudfrooij 2002; Allen \etal 2006), 
 this X-ray cavity is not found coincident with radio synchrotron emission
 (see \S\ref{sec:bubbles} and Fig. \ref{fig:centralbubbles}). Thus it
 is a 'ghost bubble' where the electron population in the radio plasma,
 expected to fill the bubble, has aged such that the synchrotron
 emission has moved to lower frequencies that are more difficult to observe 
 (Birzan \etal 2004). Diffuse X-ray emission is also observed from the
neighboring galaxy NGC\,5845. 

\subsection{Gas Densities: Evidence for Sloshing}
\label{sec:slosh}

\begin{figure}[t]
\begin{center}
\includegraphics[height=3in,width=1.95in,angle=270]{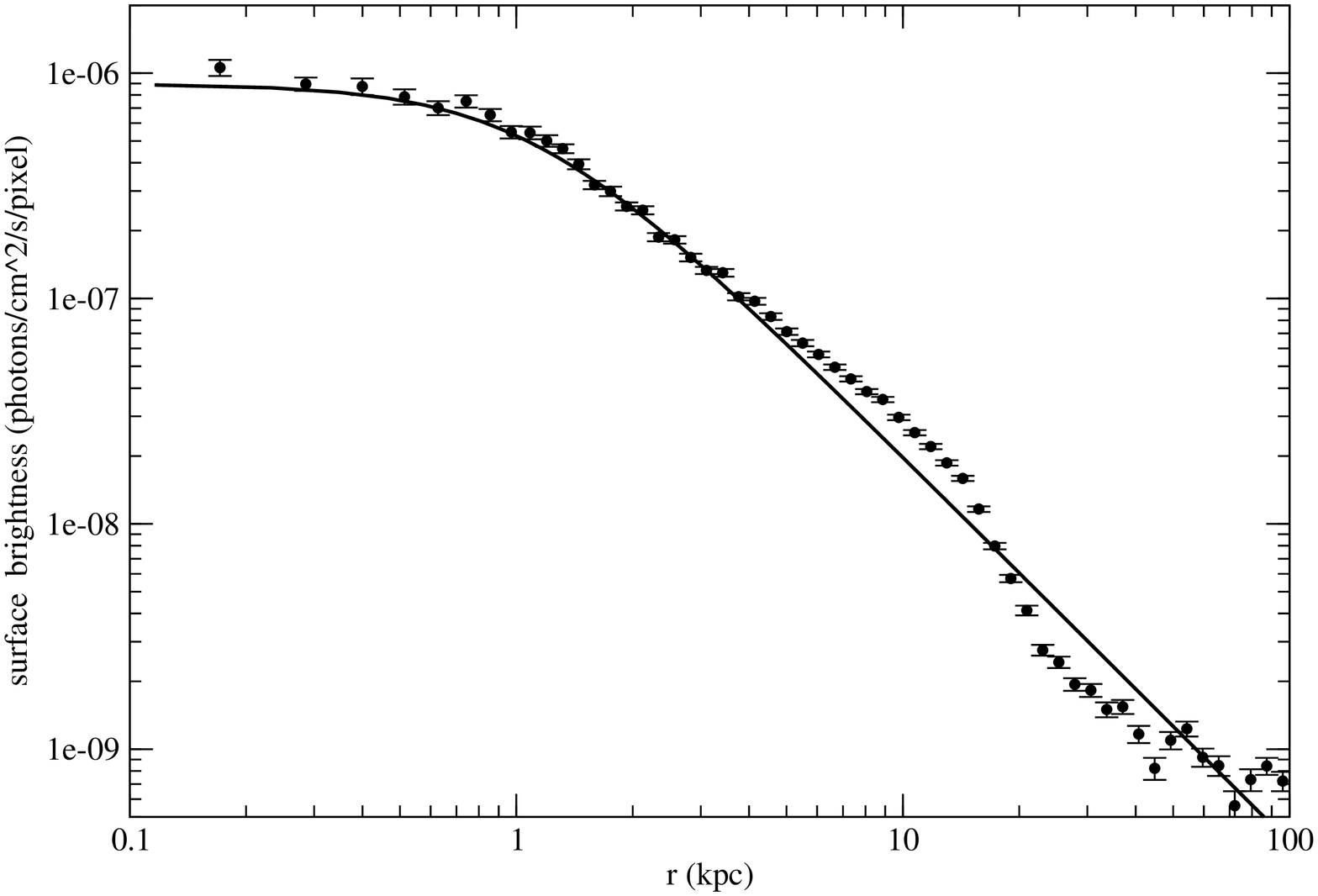}
\includegraphics[height=3.0in,width=2.14in,angle=270]{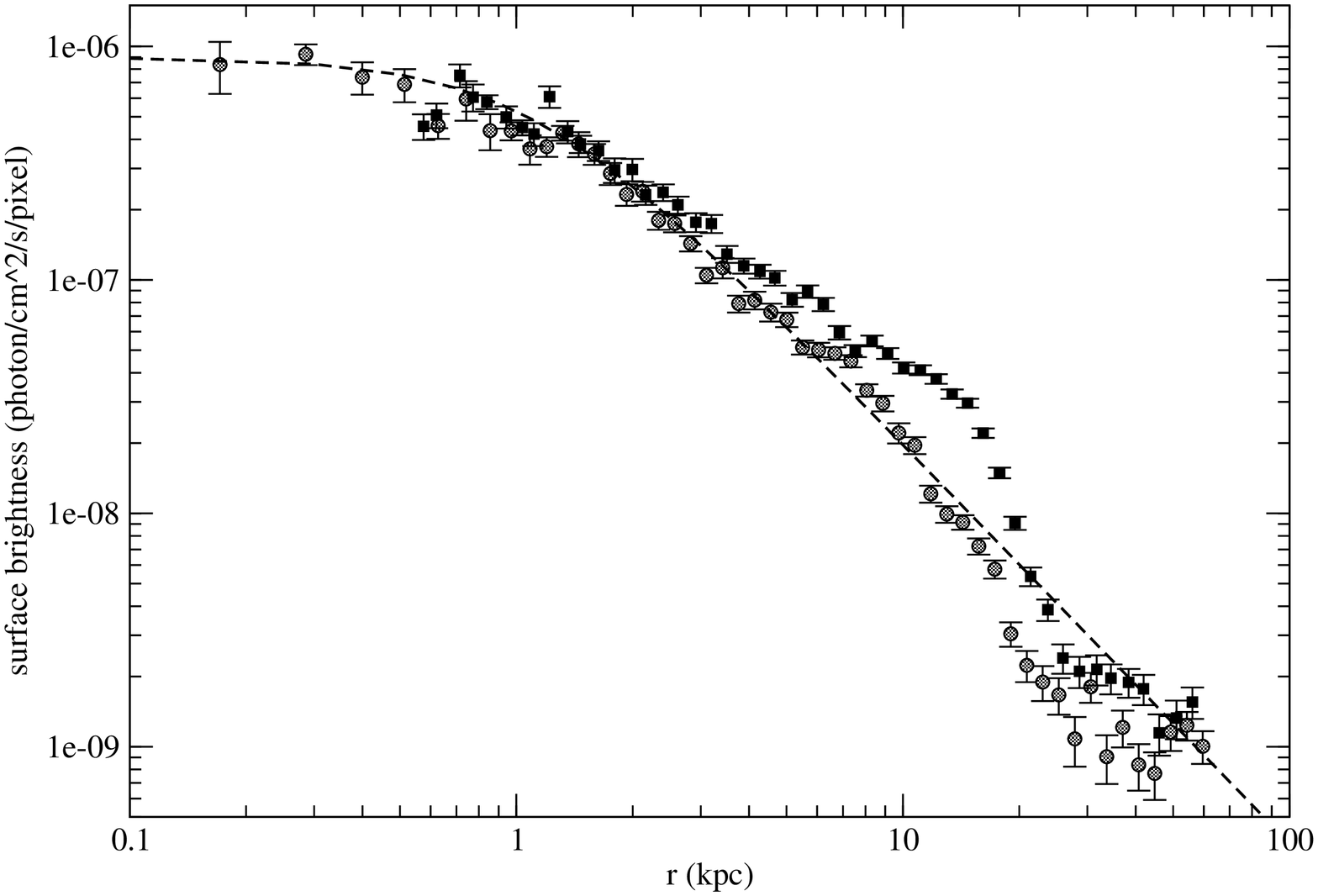}
\caption{{\footnotesize ({\it upper}) Mean radial surface brightness 
profile of NGC5846 in the $0.5-2.0$\,keV energy band.  The black
    curve is the best fit single $\beta$-model with $\beta= 0.452$, 
core radius $r_c=1.08$\,kpc ($9\farcs3$) and central surface brightness 
$s_0=8.95 \times 10^{-7}$\photflux. 
$1\,{\rm pixel} = 1\farcs0 \times 1\farcs0$. ({\it lower})  
$0.5-2.0$\,keV surface brightness radial profiles measured 
from $123^\circ$ to $180^\circ$ to the northeast (black squares)
and from $268^\circ$ to $338^\circ$ to the southwest (gray circles) compared to
the mean beta model shown in the left panel (dashed line). All
angles are measured counterclockwise from west.   
}
} 
\label{fig:meanprof}
\end{center}
\end{figure}

To investigate these  asymmetries in more detail, we use SHERPA to 
fit a single component $\beta$-model to the mean emission from
NGC\,5846 and then study deviations from this mean. 
In the upper panel of Figure \ref{fig:meanprof}, we show the  $0.5-2$\,keV 
azimuthally, averaged radial surface brightness profile, centered on
the nucleus of NGC\,5846 superposed with the  
best-fit single $\beta$-model (solid line). We find that within
$4$\,kpc the surface brightness profile is best fit by 
a single $\beta$-model with parameters $\beta= 0.452$, 
core radius $r_c=1.08$\,kpc ($9\farcs3$), and central surface 
brightness $s_0=8.95 \times 10^{-7}$\photflux, corresponding to a
central electron density of $n_0=0.091$\cmc.
While $\beta$ is in reasonable agreement with
previous {\it ROSAT} results ($\beta \sim 0.5$), we find a 
factor $\sim 1.6-2$ smaller core radius and 
infer a factor $2$ lower mean central electron density compared to
previous results (Trinchieri \etal 1997; Osmond \& Ponman 2004).  These 
discrepancies are likely a reflection of the different angular
resolution of the detectors and the failure of a 
single $\beta$-model to  describe the asymmetric surface brightness 
distribution.

\begin{figure}[t]
\begin{center}
\includegraphics[width=3.0in]{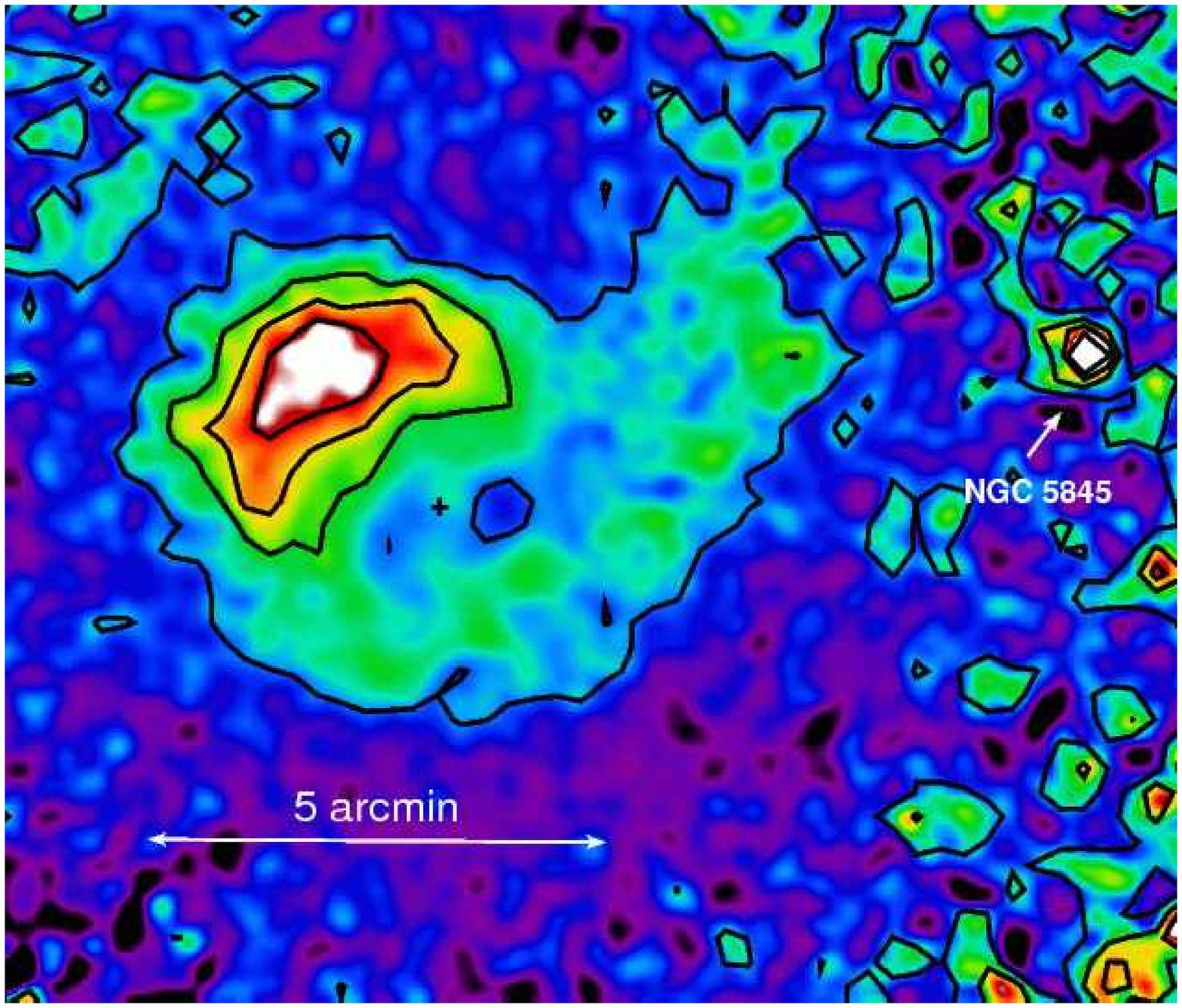}
\includegraphics[width=3.0in]{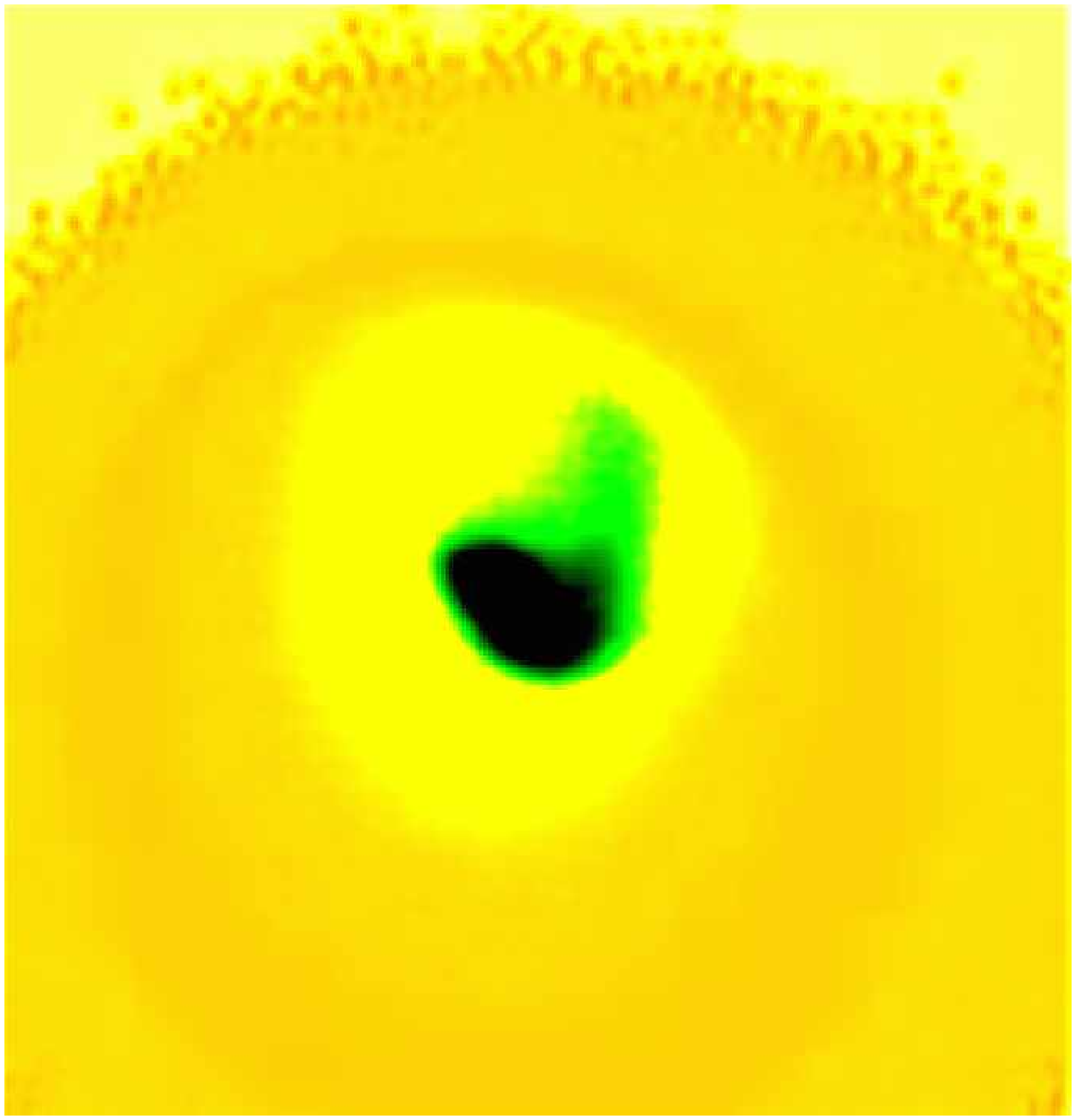}
\caption{{\footnotesize  
({\it upper}) 
Fractional difference map ($(Im-Mod)/Mod$) between the $0.5-2$\,keV
X-ray surface brightness image of the 
NGC\,5846 group ($Im$) and the best-fit $1$-dimensional 
$\beta$-model ($Mod$) for the data (see
Fig. \protect\ref{fig:meanprof}). Both $Im$ and $Mod$ were  
smoothed with a $7\farcs5$ Gaussian kernel. The nucleus of NGC\,5846 is
denoted by a black cross and the $\sim 30''$ ($3.4$\,kpc) diameter 
X-ray `ghost' cavity is outlined by the circular contour to its 
west.  $1\,{\rm arcmin} = 7$\,kpc and the panel size is $90.5$\,kpc.
Note that the linear sequence of blobs in the spiral tail are
artifacts due to the  ACIS-I chip gap. 
({\it lower}) The predicted X-ray emission from a simulated galaxy
cluster merger, that shows similar structure caused by 'sloshing'. The
image was adapted from Fig. 19, Ascasibar \& Markevitch (2006), and
the panel size is $250$\,kpc.
}}
\label{fig:ratiomap}
\end{center}
\end{figure}
In Figure \ref{fig:ratiomap} we present the fractional difference map
between the observed data and the best-fit $1$-dimensional
$\beta$-model for the group gas emission (i.e. left panel
of Fig. \ref{fig:meanprof}). We see clearly the characteristic 
spiral pattern in surface brightness (and thus density) generated 
by 'sloshing'. Gas, initially displaced by 
the gravitational attraction and/or hydrodynamical forces 
from the off-axis encounter, is 
set into oscillation as the ram-pressure suddenly decreases and the 
gravitational potential energy and 
angular momentum imparted to the galaxy gas by the close encounter 
causes the higher density core gas to rebound and overshoot the center 
of the group dark matter potential (black cross in
Fig. \ref{fig:ratiomap} east of the X-ray `ghost' cavity) and also spiral 
outward (see, e.g. simulations by Ascasibar \& Markevitch 2006; ZuHone
\etal 2010). The faint northwest 
extension, identified in Figure \ref{fig:n5846full} and seen clearly 
in Figure \ref{fig:ratiomap}, is the tail of this spiral structure, as 
denser core gas is pushed to larger radii in response to the
encounter. This tail-like structure is similar to that first found in the X-ray
surface brightness map of the NGC\,5098 group (Randall \etal 2009). 
In the lower panel of Figure \ref{fig:ratiomap}, we show the
simulated X-ray emission from a cluster merger adapted from Figure 19
of Ascasibar \& Markevitch (2006), showing similar spiral density
structure and a tail-like density enhancement caused by the 
non-hydrostatic bulk motions (sloshing) of gas in the cluster core. 
Since the simulation was for a more massive cluster merger we do not
expect the scales to be the same. However,
the similarity of our observation of sloshing in groups with features
in  simulations of galaxy cluster mergers demonstrates that such surface
brightness features are common to off-axis galaxy, group 
and cluster encounters over a wide range of mass scales (see,
e.g., Fig. 1 and Fig. 9 in Johnson \etal 2010 for galaxy cluster Abell 1644; 
 Fabian \etal 2011 for the Perseus cluster; Randall \etal 2009 for 
 the NGC\,5098 galaxy group; Machacek \etal 2010 for the NGC\,6868 group). 

However, as also seen in Figure 19 of Acasibar \& Markevitch (2006),
the projected surface brightness and corresponding temperature 
features are highly dependent on the viewing angle between the plane
of the encounter and the observer. If one ignores the NGC\,5846 
ghost bubble and the faint northwestern surface brightness 
tail, the image of NGC\,5846 in Fig. \ref{fig:n5846full}  
shows two ellipsoidal  edges to the southwest and, although the inner 
features  are less pronounced, at least two edges to the northeast. 
Such multiple surface brightness edges 
are also characteristic of gas sloshing (Markevitch \etal
2001) and are similar to the y projection 
of simulated cluster mergers in the right panel of Figure 19 from  
Ascasibar \& Markevitch (2006). The orientation of the plane of the 
encounter, that caused gas sloshing in NGC\,5846, with respect to the 
observer's line of sight likely lies between the $z$ and $y$ simulation 
projections, such that the sloshing spiral is tilted with respect to
the plane of the sky.   
This is consistent with the identification of the 
perturber as the disturbed spiral galaxy NGC\,5850, whose relative 
radial velocity with respect to NGC\,5846 is large ($\sim 770$\kms). 
In this scenario, NGC\,5850 would have approached in front of
NGC\,5846 from the west, measured relative to the elliptical galaxy, 
 and is now receding behind NGC\,5846 to the east.

\subsection {Gas Temperatures}
\label{sec:temps}

We need to understand the temperature structure of the gas to 
understand the thermodynamic state of the gas and determine relative 
gas pressures and velocities.  We first extract a mean spectrum 
for NGC\,5846 from OBSID 7923 using a $168''$ circular region centered at 
NGC5846's nucleus (RA, DEC $= 15^h06^m29.284^s$,
$+01^{\circ}36'20\farcs25$) and local background
from a circular annulus also centered at the galaxy nucleus, with 
inner, outer radii of $300''$, $378''$, respectively.   
We model the spectrum using a VAPEC model with fixed Galactic 
absorption ($4.24 \times 10^{20}$\cmc), and allow 
\begin{figure}[t]
\begin{center}
\includegraphics[width=3.0in]{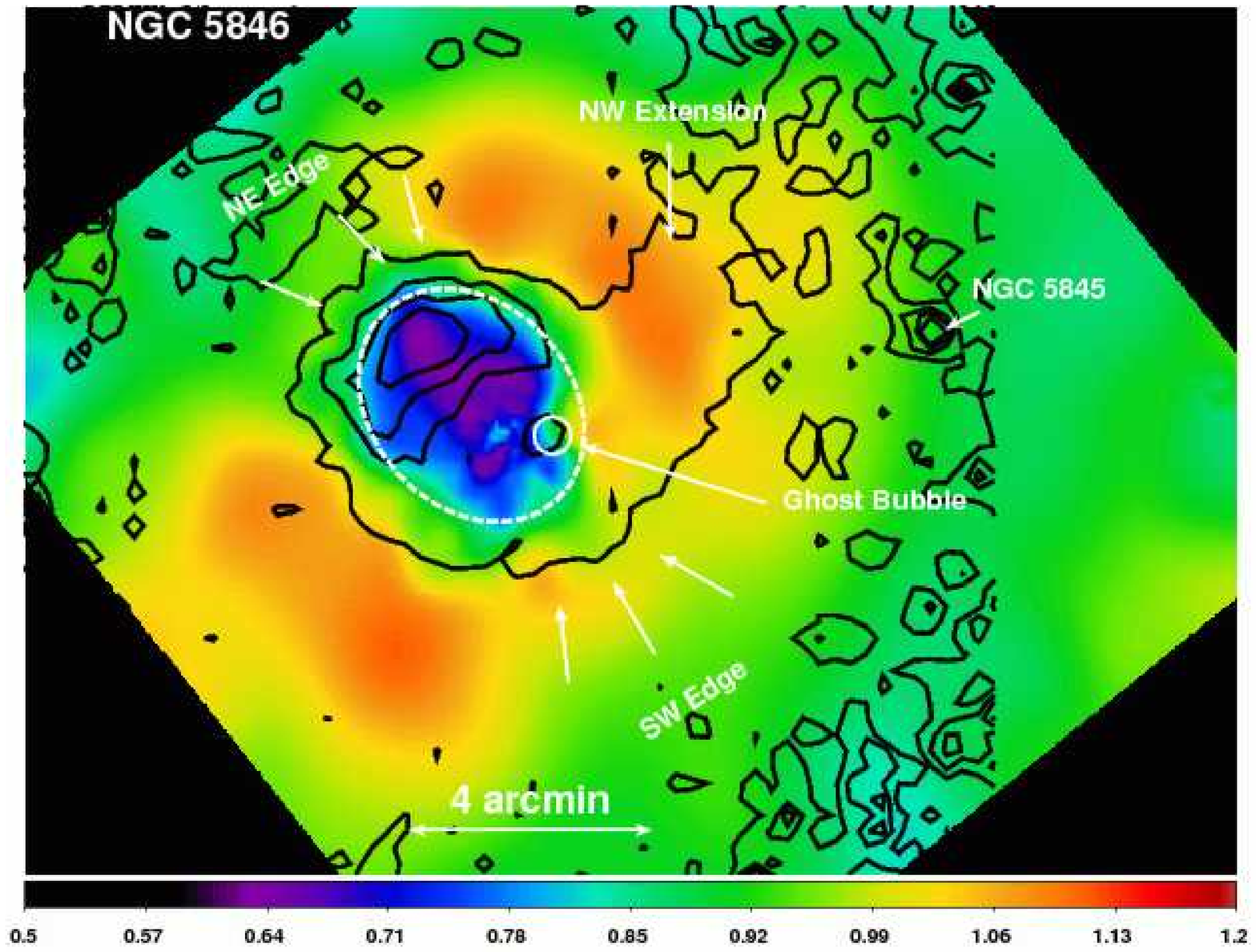}
\includegraphics[width=3.0in]{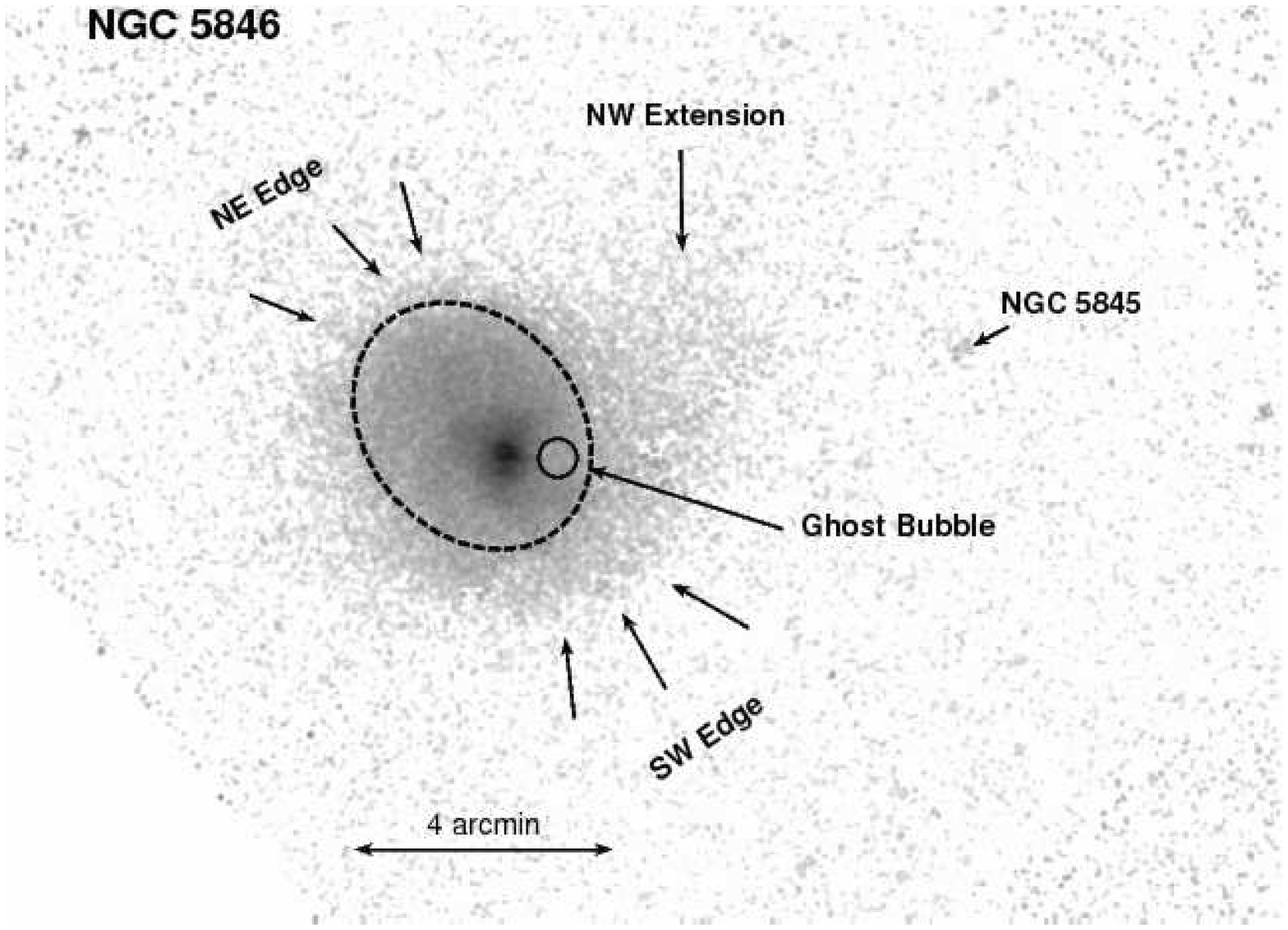}
\caption{{\footnotesize  ({\it upper)} Temperature map constructed
  using the mean energy in the $0.7-1.2$\,keV energy band (Fe peak)
  calibrated to temperature using an absorbed single APEC model with 
  Galactic absorption and abundance fixed to $0.3\,\Zs$. Black lines 
  are contours from the fractional difference
  surface brightness image (see Fig. \protect\ref{fig:ratiomap}),
  while the dashed ellipse traces the outer ($19.8$\,kpc) 
  northeastern and inner ($11$\,kpc)
  southwestern surface brightness edges shown in the lower panel. 
({\it lower}) $0.75-2$\,keV background subtracted, exposure corrected 
merged image of NGC\,5846, with the dashed ellipse tracing the 
northeastern and inner southwestern  edges. Both the temperature map
  and surface brightness maps are matched in WCS coordinates. Note
  that in both panels, the nucleus of NGC\,5846 
 is located east of the circular region denoting the ghost bubble, and 
 is significantly offset to the southwest from the center of the  ellipse.}
}
\label{fig:tmapall}
\end{center}
\end{figure}
O, Mg, Si, and Fe abundances to vary independently. 
All other abundances are fixed at a 
common fixed value $A=0.5\,\Zs$. We fit the spectrum over the
$0.5-2.0$\,keV energy range to optimize the signal relative to background.
Although we find a mean temperature of $0.65 \pm 0.01$\,keV, and 
O, Mg, Si,and Fe abundances of $0.01$, $0.34$, $0.45$, 
and $0.26\,\Zs$, respectively, this single temperature VAPEC model
does not provide a statistically 
acceptable fit to the data ($\chi^2/{\rm dof} =218/95$), 
suggesting the presence of multi-temperature gas and strong 
temperature gradients.
 
\subsubsection{Temperature Maps}
\label{sec:tmap} 
We investigate these temperature gradients by constructing a
temperature map for the region.   
From the azimuthally averaged global temperature, we know that the
temperature of the ISM  and group IGM gas are both cool. 
The thermal spectrum  for gas temperatures $\lesssim 1$\,keV
is dominated by line emission from the the FeL complex. 
For the ACIS detector, there
is a nearly linear correspondence between the mean energy measured 
in the FeL peak ($0.7 \lesssim E \lesssim 1.2$\,keV) and the
temperature of the gas, as measured by a single temperature APEC model 
(see, e.g. David \etal 2009; Machacek \etal 2010, Fig. 9). 
 We exploit
this correspondence to construct a temperature map for the core of the 
NGC\,5846 group.  
We first calibrate the 
temperature of the gas with the observed mean energy in the
$0.7-1.2$\,keV bandpass by fitting a series of simulated absorbed APEC 
model spectra, generated from the spectral response files for our 
observation, for which the 
mean energy in the Fe  peak increases from $0.3$\,keV to $2$\,keV in 
$0.05$\,keV intervals. We assume Galactic absorption and 
fix the abundance at an average value of $0.3\,\Zs$. 
Although this is higher than the $\sim 0.1\,\Zs$
metallicities observed near $r_{500}$ in the intra-group gas in 
other cool groups (Rasmussen \& Ponman 2007), and lower than the mean
metallicity $\sim 0.5\,\Zs$ for ISM gas in elliptical galaxies, the 
temperature is insensitive to changes in the abundance over 
this range, so that this does not significantly affect our results. 
We average the energy over regions determined by adaptively smoothing the 
$0.7 - 1.2$\,keV {\it Chandra} image of the NGC\,5846 group and use 
the derived calibration curve to associate the mean energy in each
region to the corresponding temperature of the diffuse gas.
Our results are shown in Figure \ref{fig:tmapall}. 

\begin{figure}[t]
\begin{center}
\includegraphics[width=3.0in]{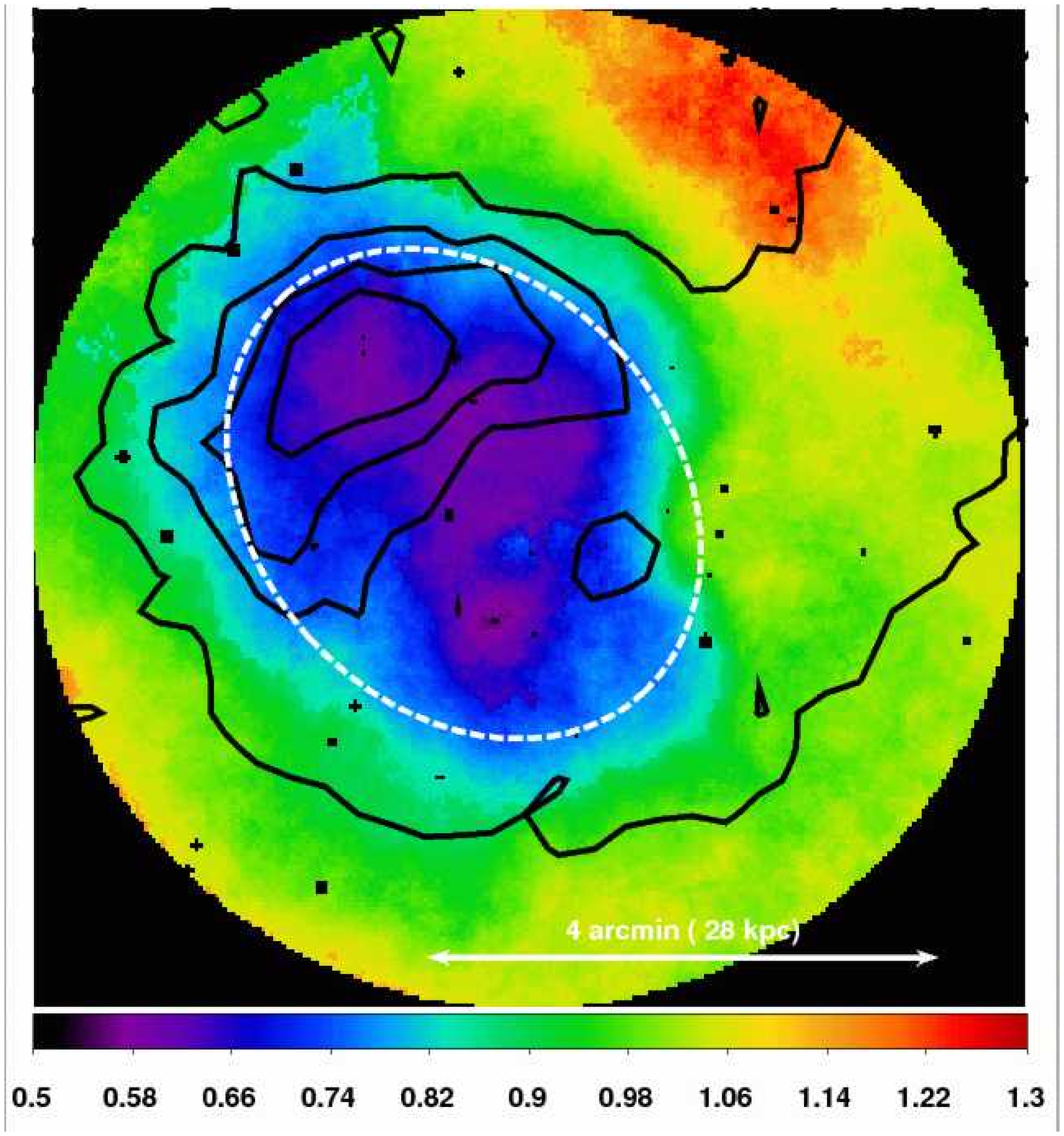}
\includegraphics[width=3.0in]{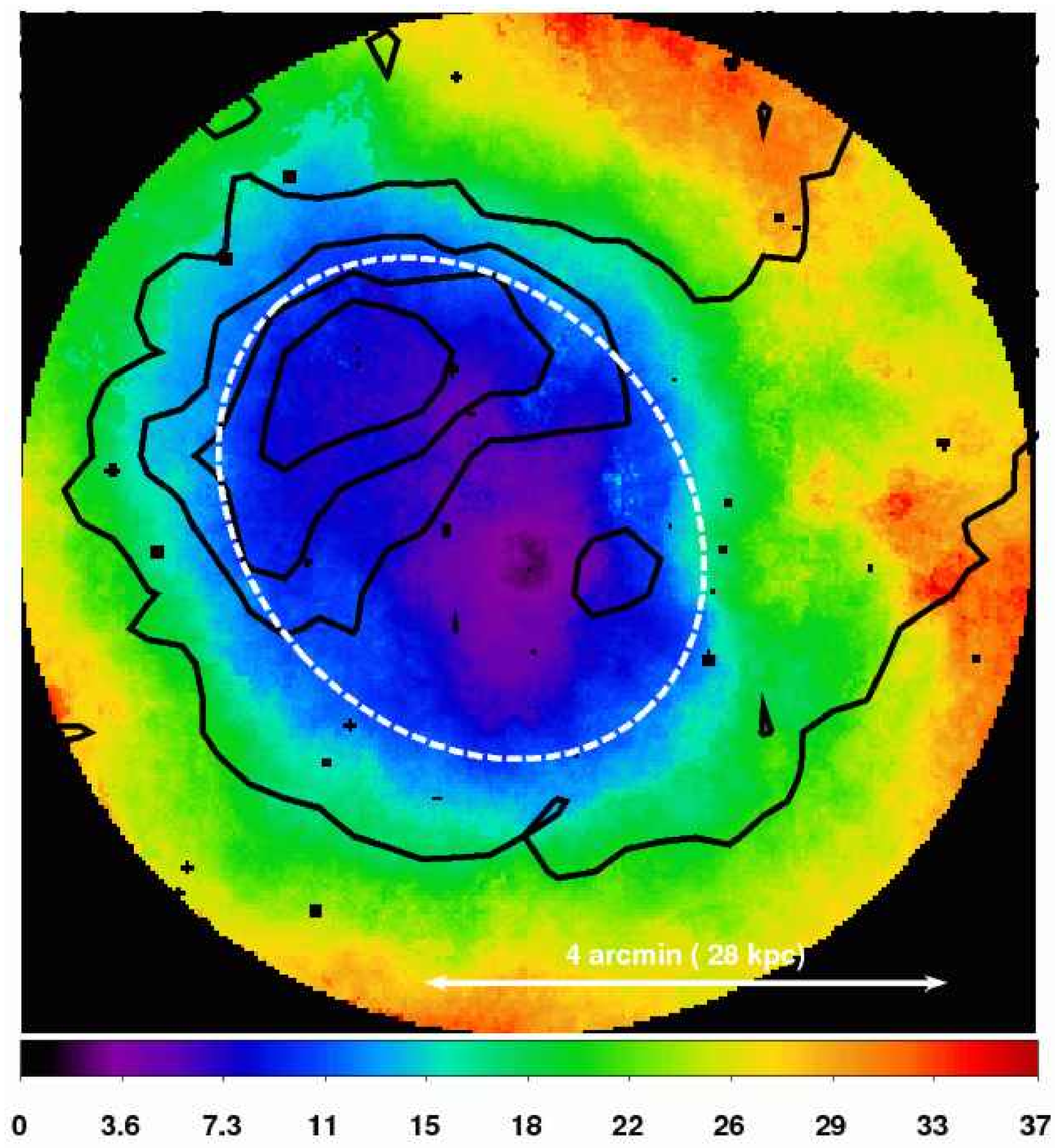}
\caption{\footnotesize {  Temperature ({\it  upper}), 
and pseudo-entropy ({\it lower})
maps of NGC\,5846 constructed using spectral 
regions, constrained to contain a minimum of $1000$ net (background
 subtracted) counts and  fit with single temperature APEC models using
 fixed Galactic absorption ($4.24 \times 10^{20}$\cmc).  
Black contours are from the fractional difference surface 
brightness image (Fig. \protect\ref{fig:ratiomap}),
  while the white dashed ellipse traces the bright northeastern and inner
  southwestern surface brightness edges, as in Fig. \protect\ref{fig:tmapall}. 
}}
\label{fig:scottmaps}
\end{center}
\end{figure}

To check these results and investigate possible temperature structure
within the central region of NGC\,5846, we construct a second
temperature map using the methods of Randall \etal (2009) to fit a single 
temperature APEC model with Galactic absorption to regions, 
containing a minimum of $1000$ counts, grown around each pixel. 
The abundance is allowed to freely vary. The results 
are shown in the top panel 
of Figure \ref{fig:scottmaps}. The temperature maps in Figures
\ref{fig:tmapall} and \ref{fig:scottmaps} are remarkably similar,
suggesting that the features identified are robust. To guide the eye
in both maps, we superpose the contours from the fractional difference 
map that shows the spiral structure in the X-ray surface brightness 
(Fig \ref{fig:ratiomap}). The dashed ellipse in both maps traces the
outer northeast and inner southwest surface brightness edges. Note
that the nucleus of the galaxy, corresponding to the X-ray bright AGN,
is offset from the center of the
ellipse. Lower temperatures are found interior to 
this ellipse and to the northeast of the nucleus, confirming that 
these edges are cold fronts. The  
gas temperature is again higher
($\sim 1$\,keV) on
the outside of the outer southwestern edge, confirming that this edge
is also a cold front.  Most of the lowest temperature 
($\sim 0.6$\,keV) gas resides northeast of the nucleus, 
along the contours 
of maximal fractional difference from the best fit $\beta$-model for the X-ray
surface brightness. A finger of slightly higher ($\sim 0.7-0.8$\,keV) 
temperature gas is drawn into the nuclear region. Cluster simulations 
 also show that sloshing bulk gas motions create 
multiple edges and may push higher 
temperature gas into the cooler cluster cores. Note, however, 
that there is no evidence for spiral structure in these temperature maps  
corresponding to the large scale spiral-like northwest extension 
seen in the X-ray surface brightness. This is in sharp contrast to the
cool, spiral-like tails observed in the NGC\,5098 (Randall \etal 2009)
and NGC\,6868 (Machacek \etal 2010) groups, and would be difficult to explain 
if the sloshing encounter occurred in the plane of the sky (Ascasibar \&
Markevitch 2006; ZuHone \etal 2010), again suggesting projection
effects are important. 

A map of 'pseudo' entropy ($s \propto T n^{-2/3}$) 
is shown in the lower panel of Figure \ref{fig:scottmaps}. 
This map is  constructed by using the best-fit APEC model
normalization $K$, where
  \begin{equation}
   K \propto n_e^2 \Big (\frac{10^{-14}V}{4 \pi [D_A(1+z)]^2} \Big ) ,
 \label{eq:napec}  
\end{equation} 
 as a proxy for $n_e^2$ for each of the spectral
regions in the temperature map. 
$D_A$ and $z$ are the angular size distance and redshift of
NGC\,5846, respectively, and for simplicity we assume a constant    
emission volume $V$ for each region. As expected, the lowest
entropy gas is found coincident with NGC\,5846's central region. Entropy 
increases smoothly with radius. However, the distribution is not azimuthally
symmetric, but again follows the off-centered ellipse (dashed line)
that traces the northeastern outer and southwestern inner edges. 
While this map is suggestive, it provides only a qualitative 
picture for entropy throughout the region. First, the uncertainties in the
abundances for spectral regions containing only $1000$ source counts are 
large. Although the model temperatures do not vary significantly with 
abundance in this range, the normalization $K$ in equation
\ref{eq:napec} does by a factor of $\gtrsim 2$ for $0.64$\,keV gas
and abundances between $0.4 -1\,\Zs$ (see, e.g. Machacek \etal 2006). 
This increases the uncertainty in the 
derived density to $\sim 40\%$ and in the derived entropy to $\sim 26\%$.
Second, the true geometries (volumes) 
of the emission regions in the disturbed gas are likely 
complex, as demonstrated by gas distributions in the cores of
simulated clusters (see, e.g. Ascasibar \& Markevitch 2006; 
ZuHone \& Markevitch 2009; ZuHone \etal 2010), and may not be 
well represented by the simple 
model used to construct the maps. Thus the qualitative features shown
in these maps need to be checked by more targeted spectral analysis
and modeling.

\subsubsection{Modeling Abundances}
\label{sec:abundances}

Since we want to constrain the relative densities between differing
spectral regions as well as possible, we choose to use the VAPEC
thermal plasma model for detailed spectral fits. To optimize signal to
noise, the fit range is restricted to the  $0.5-2$\,keV band, 
where emission from cool $\lesssim 1$\,kev gas dominates.
Guided by the temperature maps in Figures \ref{fig:tmapall} and 
\ref{fig:scottmaps}, we extract the spectrum for an X-ray luminous 
region expected to be homogeneous in temperature, defined as the sum
of two circular regions with radius of
$39''$ centered at ($15^h06^m29.85^s$, $+01^\circ37'14\farcs89$) and 
$31\farcs4$ centered at ($15^h06^m34.445^s$,
$+01^\circ37'38\farcs50$), respectively. We model
the spectrum of this region using an absorbed  
 VAPEC model with fixed Galactic absorption, but with the abundances 
for O, Mg, Si, and Fe, expected to be important for low temperature
gas, free to vary. The remaining abundances for elements that do not 
have strong emission features in the $0.5-2$\,keV energy band are
fixed at $0.5\,\Zs$.  We find a temperature of $0.60 \pm 0.01$\,keV
and O, Mg, Si, and Fe abundances of $0.24^{+0.17}_{-0.12}\,\Zs$, 
$0.53^{+0.17}_{-0.13}\,\Zs$, $0.72^{+0.21}_{-0.16}\,\Zs$, and 
$0.49^{+0.11}_{-0.08}\,\Zs$, respectively ($\chi^2/{\rm dof} = 79.8/79$).
We then fix the metal abundances at these values in our spectral
models for the less luminous features.
 
\begin{figure}[t]
\begin{center}
\includegraphics[height=3.0in,angle=270]{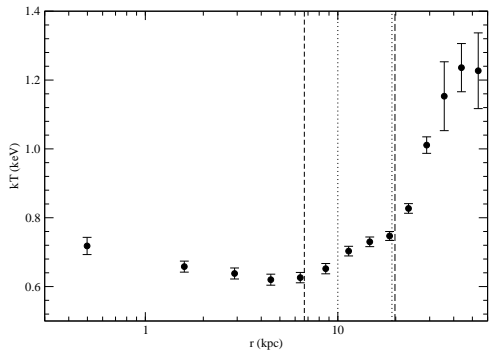}
\caption{{\footnotesize Projected, radial temperature profile of
  the NGC\,5846 group. The spectra for logarithmically increasing
  bin widths are modeled using an absorbed 
  VAPEC plasma model with fixed Galactic absorption 
(see Table \protect\ref{tab:tprof}). Vertical lines denote the
  positions of edges to the northeast (dashed) and southwest (dotted). 
 (see Figs. \protect\ref{fig:n5846full} and \protect\ref{fig:tmapall} 
   and \protect\S\ref{sec:densjump}).}
} 
\label{fig:tprof}
\end{center}
\end{figure}

\subsubsection{Azimuthally Averaged Temperature Profile}
\label{sec:tprof}

In Figure \ref{fig:tprof}, 
 we show the azimuthally averaged temperature
profile of NGC\,5846 to a maximum radius of $40$\,kpc using $14$
circular annuli, with logarithmically increasing bin width ranging from
 $6''$ ($0.7$\,kpc) for the innermost bin to $64\farcs2$ ($7.45$\,kpc)
 for the outermost bin (see Table \ref{tab:tprof}). The local 
background for all bins was chosen to be a  $3'$ radius circular 
region, centered at RA, Dec of $15^h05^m52.374^s$,
 $+01^\circ35'43\farcs9$, placing the background region 
 as far as possible ($9\farcm2$) from
 NGC\,5846 to minimize the contribution from 
group emission, while still being completely contained  on the 
ACIS-I detector. 

We model the spectra  with a single  absorbed VAPEC model,  Galactic 
absorption and abundances fixed as in \S\ref{sec:abundances} and fit 
the spectrum over the $0.5-2$\,keV range. Our results are listed in 
Table \ref{tab:tprof}. A spectral fit is deemed statistically acceptable if the
probability of getting a $\chi^2$ as large or larger than observed
if the model is correct, denoted the `null' parameter, is
${\rm null} > 0.05$.
Although the  single temperature VAPEC model 
 provides a reasonable description of the spectra outside the central 
$1.5$\,kpc of the galaxy, the $\chi^2/{\rm dof}$ is only marginally acceptable 
(null $\sim 0.01-0.02$) for bins with mean radii of $8$\,kpc,
 $16$\,kpc and $25$\,kpc, and the residuals in those bins suggest the 
presence of  multi-temperature gas (Buote 2000). This is not surprising 
 given the elliptical, off-centered distribution of temperatures 
shown in the  temperature maps. Within  $10$\,kpc, the gas
temperature is roughly isothermal with temperatures of $0.62 -
0.65$\,keV. The mean temperature rises modestly to $\sim 0.75$\,keV at 
 $\sim 10$\,kpc and then rises steeply to $\gtrsim 1$\,keV at radii 
beyond $20$\,kpc, consistent with {\it ROSAT} and {\it XMM-Newton} results 
(Trinchieri \etal 1997; Finoguenov \etal 1999; Nagino \& Matsushita 2009).
The temperature increases for increasing radii at $\sim 8$\,kpc and
$20$\,kpc are approximately coincident with the positions of the inner
and outer edges seen in the surface brightness profiles
(Fig. \ref{fig:edgefits}). In the central $0.7$\,kpc, a single
temperature VAPEC model, with best fit temperature $0.72 \pm
0.02$\,keV ($\chi^2/{\rm dof}= 76/48$, ${\rm null} = 0.006$) 
is not a good description of the data, with
the see-saw pattern in the residuals indicating the presence of
multi-phase gas. A two temperature VAPEC model with abundances fixed
as in \S\ref{sec:abundances} and temperatures 
$0.60 ^{+0.04}_{-0.06}$\,keV and $1.3^{+0.34}_{-0.18}$\,keV 
($\chi^2/{\rm dof}= 31/46$, ${\rm null} = 0.95$)is a much better
description of the data.  The single temperature VAPEC model also fails
to describe the spectrum in the $0.7 - 1.4$\,kpc annulus. This central 
region $r \lesssim 1.5$\,kpc) is complex, showing X-ray
cavities (bubbles)  filled with radio plasma from recent AGN activity,
and X-ray bright knots as well as H$\alpha$ emission 
(Fig. \ref{fig:centralmontage} and \S\ref{sec:bubbles}).
The presence of higher temperature gas in the spectral model for the 
central region of NGC\,5846 is unusual for group dominant 
elliptical galaxies, whose bright cores tend to be cool. 
This spectral component may be an artifact of applying a simple
spectral model to multi-phase gas in the central region of NGC\,5846, or 
 indicate the presence of higher temperature gas 
that has been heated by the recent passage of a shock.   

\subsection{Modeling Gas Properties Across the Surface Brightness Edges}
\label{sec:edges}

\subsubsection{Gas Densities}
\label{sec:densjump}
   
To study the northeastern edges quantitatively, we identify a bounding 
ellipse centered at NGC\,5846's nucleus with semi-major (-minor) axes of 
$2\farcm49$ ($2\farcm16$) and position angle $42.3^\circ$,
that traces the outer sharp surface brightness discontinuity in the 
angular sector extending from $123^\circ$ to $180^\circ$. We 
construct the surface brightness profile, in that angular 
sector,  using concentric elliptical annuli 
of increasing (decreasing) logarithmic width for radii
outside (inside) the bounding  ellipse. To the southwest, we construct 
the radial surface brightness profile using  
concentric circular annuli, centered again on NGC\,5846 and
with logarithmically increasing width, constrained to lie in the
angular sector extending from $268^\circ$ to $338^\circ$.  
In the right panel of Figure \ref{fig:meanprof}, we compare the 
mean $\beta$-model (dashed line) with these surface brightness
profiles, 
confirming the qualitative evidence for multiple edges seen in 
Figure \ref{fig:n5846full}. Not only is the
surface brightness and thus gas density much higher at $\lesssim 20$\,kpc 
inside the outer northeast edge, as also seen in ROSAT HRI images 
(Trinchieri \etal 1997), but the edge is sharper and deeper than that to the
southwest, suggesting that the ISM gas distribution has been displaced 
significantly to the northeast. The surface brightness profile to the
southwest is a factor $2$ fainter at $\sim 10$\,kpc compared to that
at the same distance from the nucleus to the northeast. 
There is a shallower discontinuity near $20$\,kpc in the southwestern 
profile than in the northeast, perhaps a remnant from a previous 
oscillation. 

\begin{figure}[t]
\begin{center}
\includegraphics[height=3.0in,angle=270]{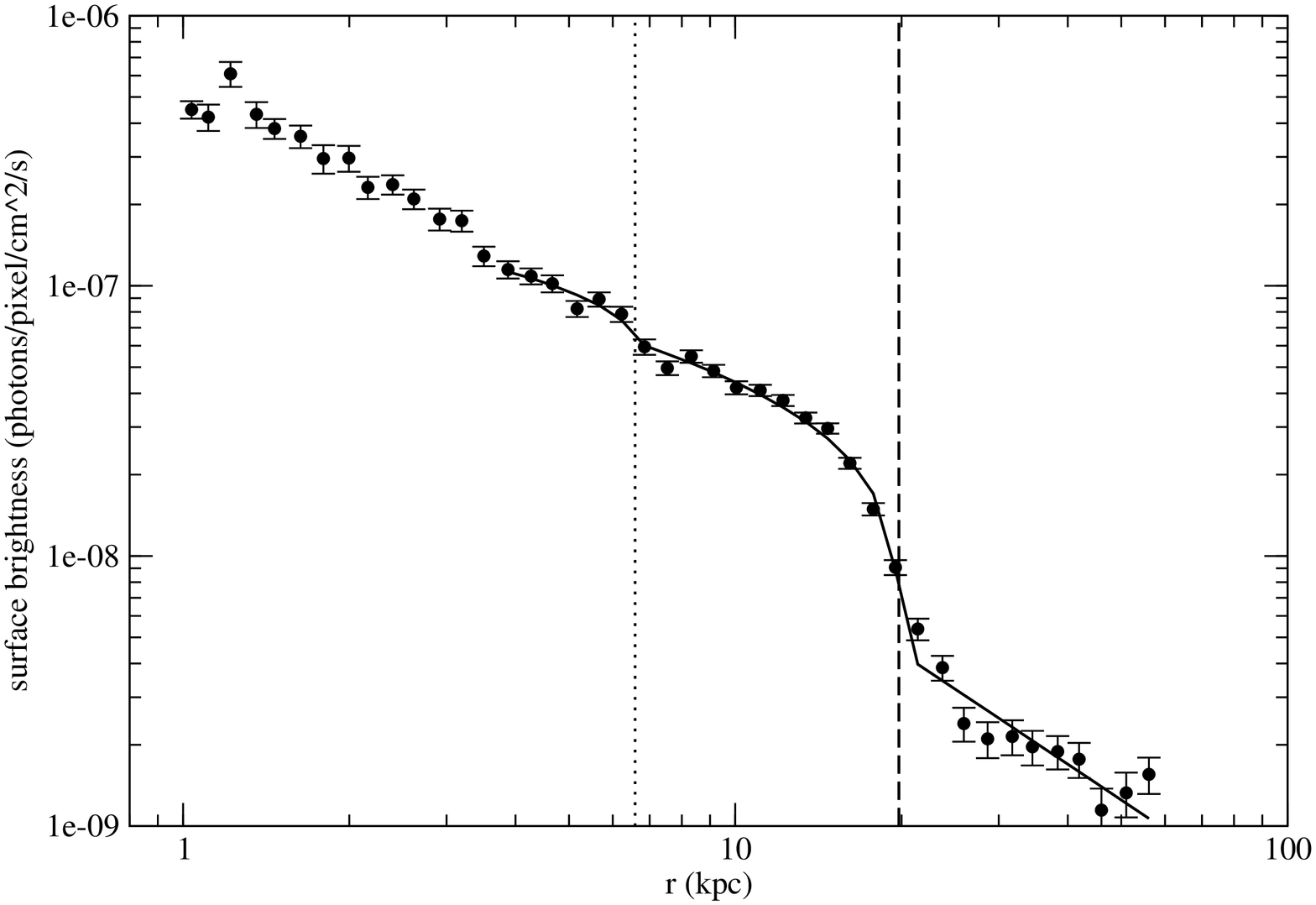}
\includegraphics[height=3.0in,angle=270]{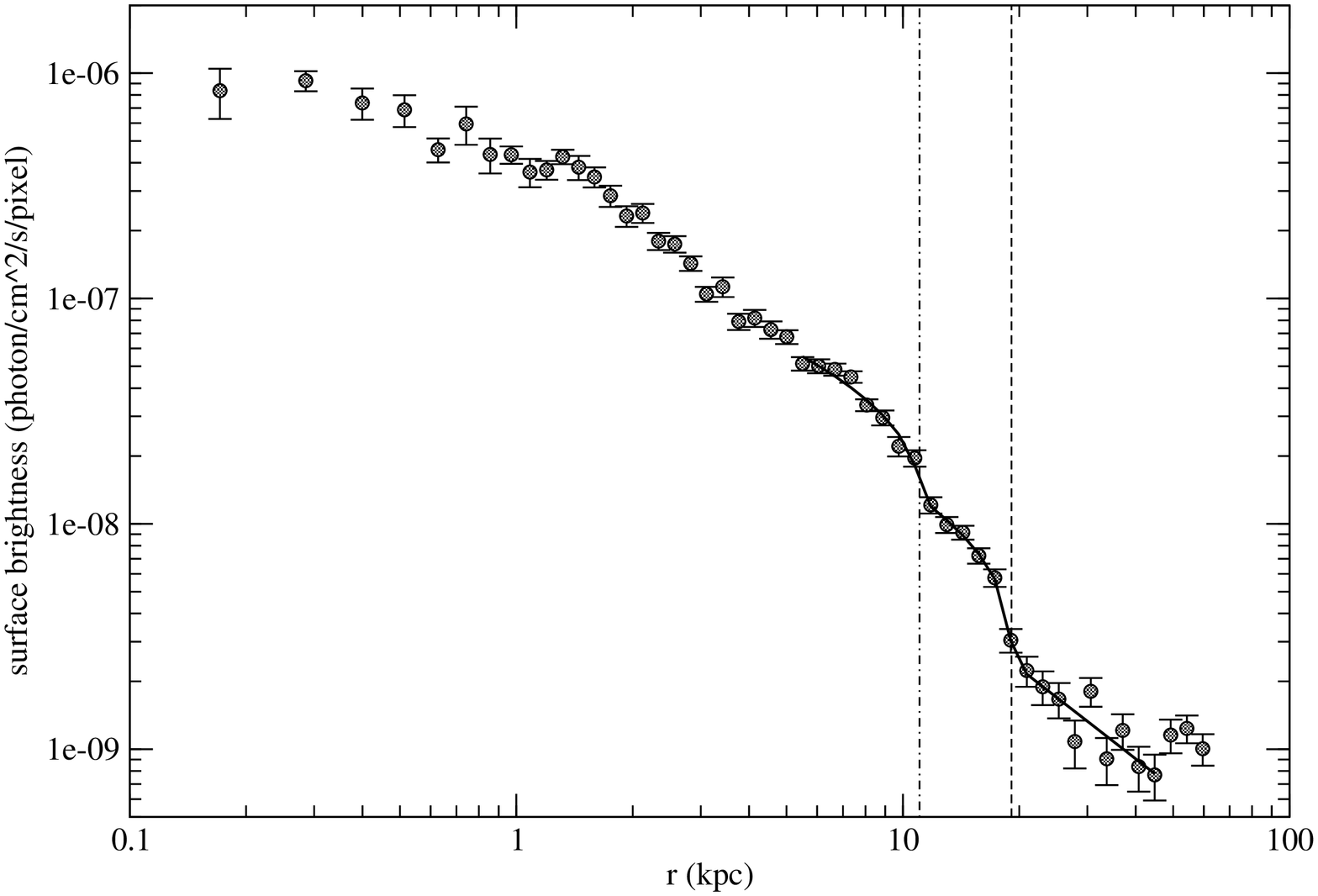}
\caption{{\footnotesize ({\it upper}) $0.5-2$\,keV surface brightness 
profile to the northeast of NGC\,5846,  taken in elliptical annuli
concentric to the bounding ellipse, centered at NGCC5846's nucleus, with 
semi-major (-minor) axes of $2\farcm49$ ($2\farcm16$) 
and position angle $42.3^\circ$ and constrained to lie in the angular
sector between $123^\circ$ to $180^\circ$. The solid line represents 
the best-fit density law model from  
equations 1 - \protect\ref{eq:densitymod}, and listed in Table
\protect\ref{tab:edgefits}. The inner (outer) edge location at 
$ \sim 6.7$\,kpc ($19.8$\,kpc) is denoted by the dotted (long dashed)
vertical line, respectively. 
({\it lower}) $0.5-2$\,keV surface
brightness profile to the southwest of NGC\,5846, taken in concentric
circular annuli constrained to lie within the angular sector between 
$268^\circ$ to $338^\circ$. The solid line is the best-fit density
model given in Table \protect\ref{tab:edgefits}. The inner (outer)
edge locations at $11.0$\,kpc ($19.1$\,kpc) are denoted by 
dot-dashed (dashed) vertical lines, respectively.}}
\label{fig:edgefits}
\end{center}
\end{figure}

Following Vikhlinin \etal (2001), we denote the edge position and
ratio of electron density across the inner (outer) edge in each sector 
as $r_1$ and $J_1$ ($r_2$ and $J_2$ ), respectively. The electron 
density $n_e$ on either side of each edge is
modeled with a power law. Thus,
\begin{eqnarray}
 n_e &= J_1 \Big  [n_0 J_2 \Big 
   ( \frac{r_1}{r_2} \Big )^{\alpha_{\rm mid}} \Big ]
   \Big ( \frac{r}{r_1})\Big )^{\alpha_1} &  ( r < r_1) \\
   &=  n_0 J_2 \Big ( \frac{r}{r_2} \Big )^{\alpha_{\rm mid}}  
    & (r_1 \leq r < r_2 )\\
   &=  n_0 \Big ( \frac{r}{r_2} \Big )^{\alpha_2} & ( r \geq r_2) 
\label{eq:densitymod}
\end{eqnarray} 
where the power law slopes of the density for radii inside the inner edge,
between the edges, and outside the outer edge are $\alpha_1$, 
$\alpha_{{\rm mid}}$, and $\alpha_2$, respectively and $n_0$ is a 
constant normalization. We then integrate the electron density model
along the line of sight and use a multivariate $\chi^2$ 
minimization scheme iteratively to fit the model parameters 
for each sector. We
first determine $\alpha_2$ and $r_2$ from a fit to the outer edge alone, 
and then fix these parameters in a fit across both edges while
allowing the remaining parameters ($J_1$, $J_2$, $r_1$, $\alpha_1$,
$\alpha_{{\rm mid}}$, $n_0$) to vary. Our results for the northeast
(NE) and southwest (SW) sectors are listed in Table \ref{tab:edgefits}
and shown as solid curved lines in the upper and lower panels of Figure 
\ref{fig:edgefits}. Edge locations are denoted by vertical lines. 
In the northeastern sector,  we find a sharp, deep
outer edge located $19.8$\,kpc from the nucleus with a gas density jump, 
assuming no strong abundance changes at the boundary, 
of $2.9^{+0.5}_{-0.4}$.
The inner edge, located at $6.6 -6.8$\,kpc, is not as well-defined and is
a factor $\sim 2$ shallower with a density jump of $1.4 \pm 0.2$. 
Similarly we find edges at $19.1$\,kpc and
$11.0$\,kpc from the nucleus in the southwest sector 
with density discontinuities of
$2.1^{+0.7}_{-0.4}$ and $1.6 \pm 0.2$,
respectively. The inner edge in the southwest sector falls close to a
chip gap on the ACIS-I detector. To test whether this might bias our
results, we fit the surface brightness profile from the same SW
sector using only data from the $30$\,ks obsid 788 observation, where 
the inner edge region of interest falls entirely on the S3 CCD, and 
found no significant differences in the fit results. We also note that
the $\beta$-model parameters derived from the outer power law slopes
($\alpha_2$) for the NE sector ($0.39^{+0.05}_{-0.04}$) and SW sector 
($0.39^{+0.06}_{-0.08}$) are in good agreement with each other and with
the $\beta = 0.45$ obtained from a single $\beta$-model fit to the
NGC\,5846 group's mean (azimuthally-averaged) radial surface 
brightness profile, shown in the upper panel of Figure \ref{fig:meanprof}.

\begin{figure}[t]
\begin{center}
\includegraphics[height=3.0in,angle=270]{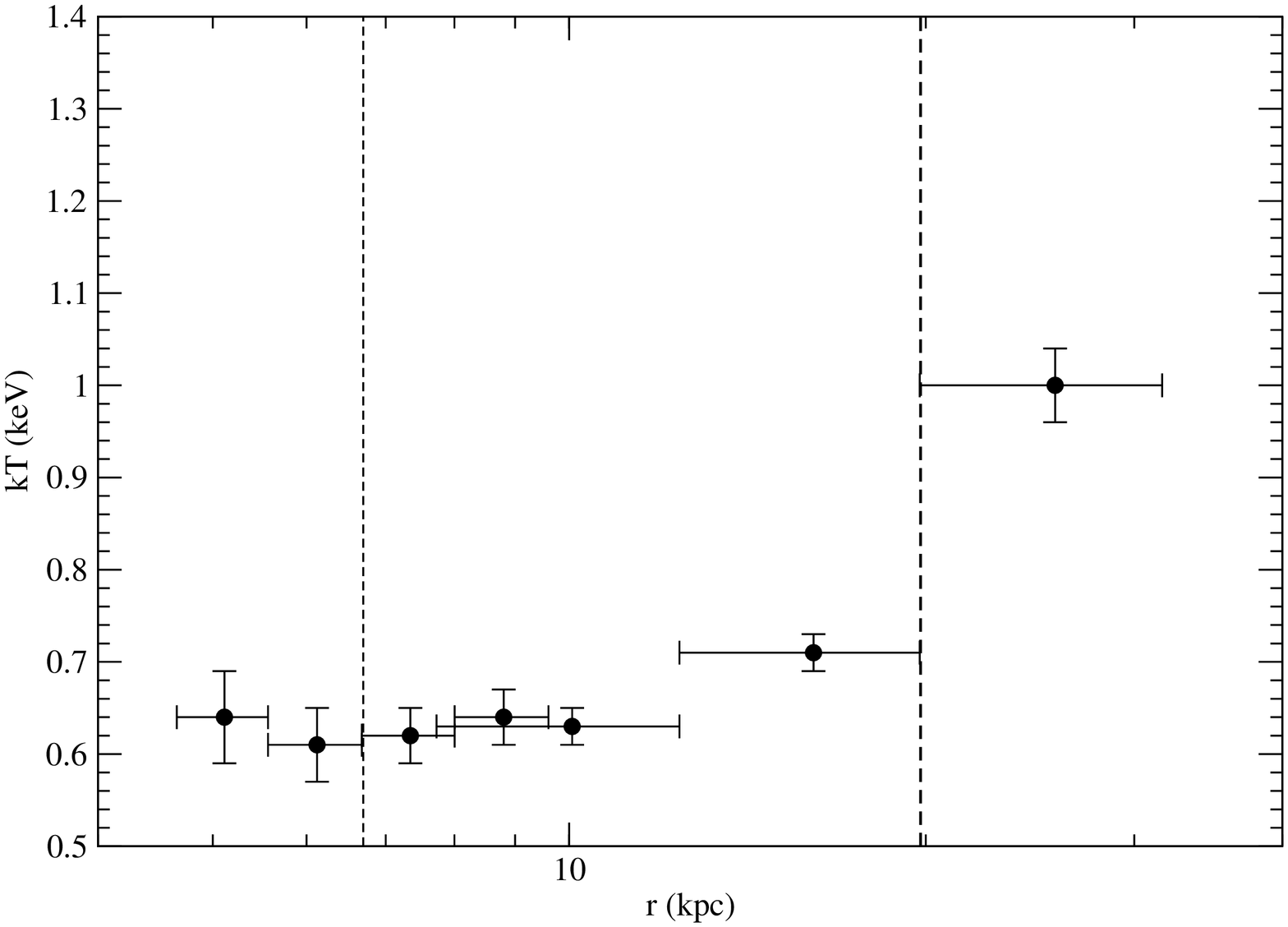}
\includegraphics[height=3.0in,angle=270]{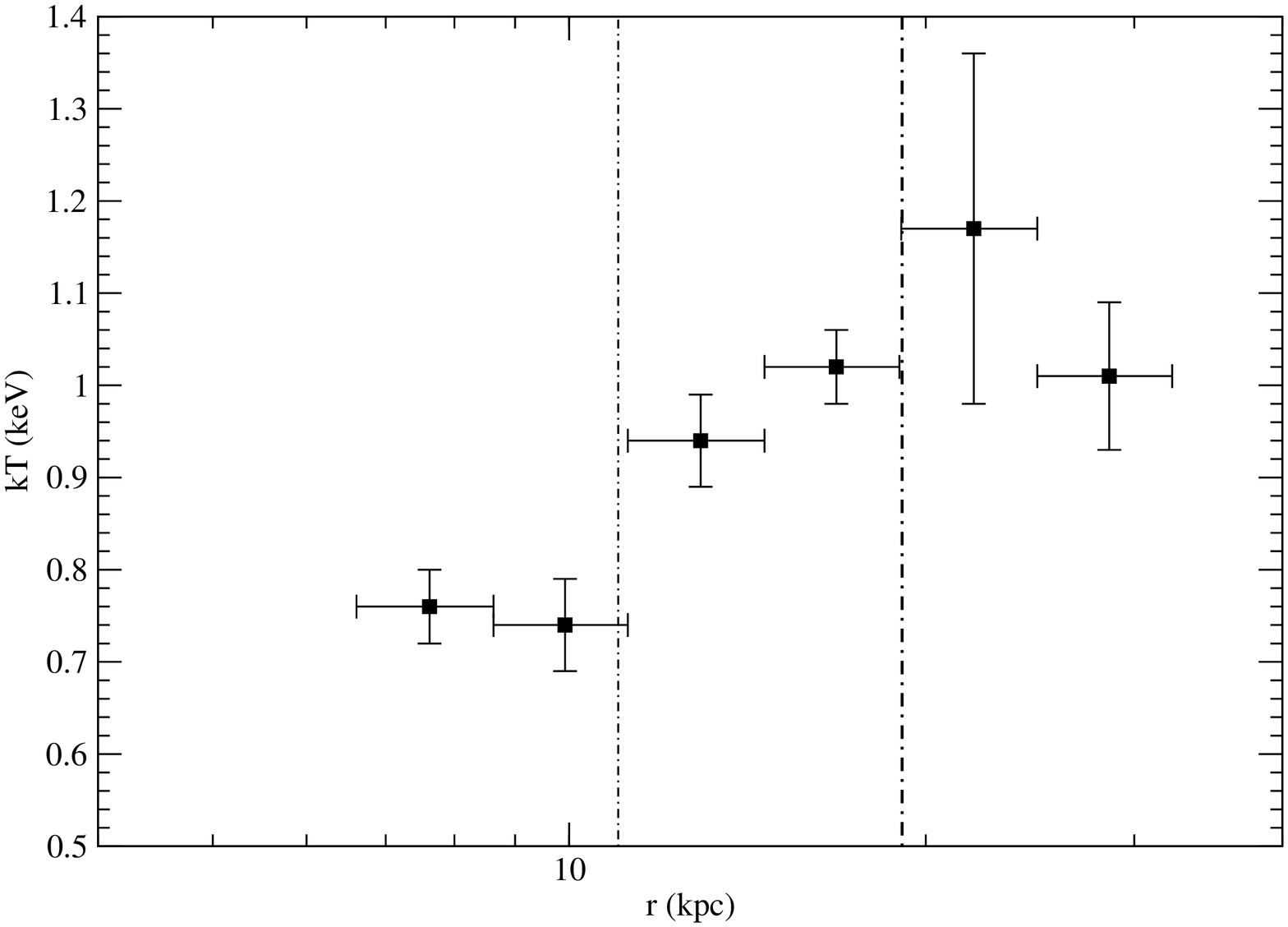}
\caption{{\footnotesize ({\it upper}) Projected, radial temperature
  profile  across the northeast surface brightness edges. Spectral
  regions  are elliptical annuli chosen to be concentric to the 
  bounding ellipse defined in Fig. \protect\ref{fig:edgefits}, and
  constrained to lie in the angular sector from $123^\circ$ to $180^\circ$ 
  (see Table \protect\ref{tab:needgespec}).
 ({\it lower})  Projected, radial temperature
  profile across the southwestern surface brightness edges. Spectral
  regions are concentric circular annuli centered on NGC\,5846's nucleus and
  constrained to lie within the angular sector from $268.2^\circ$ to
  $338^\circ$ (See Table \protect\ref{tab:swedgespec}).   
  For each case vertical lines denote the edge
  positions, and the spectra are modeled using an absorbed 
  VAPEC plasma model with fixed Galactic absorption and abundances
  as in \S\protect\ref{sec:abundances} and Table \protect\ref{tab:tprof}. 
}} 
\label{fig:edgetprof}
\end{center}
\end{figure}

\subsubsection{Gas Temperatures, Pressures and Velocities}
\label{sec:edgetemp}

To better understand the temperature changes across the surface
brightness edges, we extract spectra from either side of the edge
using regions concentric with the edge morphology. For the northeastern edges, 
we extract spectra from elliptical annular regions 
concentric to the same bounding ellipse that was used to generate the 
northeastern sector surface brightness profiles and constrained to lie
in the same angular sector (see \S\ref{sec:slosh} and
Fig. \ref{fig:edgefits}). Regions NEO-1
(NEO0) fall just inside (outside) the outer northeastern edge
at $19.8$\,kpc. NEI-1 (NEI0) lie inside
(outside) the inner northeastern edge at $\sim 6.7$\,kpc.  
Similarly for the southwestern edges, we use concentric circular
annuli, centered on NGC5846's nucleus and 
constrained to lie in the same sector used to construct the 
southwestern surface brightness profile, to measure the gas
temperatures on either side of each edge.
Regions SWO-1(SWO0) fall just inside (outside) the outer  
southwestern  edge at $19.1$\,kpc, and SWI-1 (SWI0) lie just inside
(outside) the inner southwestern edge at $11.0$\,kpc, respectively.  
Backgrounds are  taken from renormalized blank sky backgrounds 
provided by the Chandra X-ray Center for the same regions, 
normalized to the observations by comparing count rates in 
the $9.0-12.0$\,keV energy band (see \S\ref{sec:obs}). 

Our results are listed in
Table \ref{tab:needgespec} for the northeastern edges and in Table 
\ref{tab:swedgespec} for the southwestern edges, and are plotted in
Figure \ref{fig:edgetprof}. 
As expected, the temperature profiles are not symmetric. Lower 
temperature gas has been pushed to larger radii to the northeast. 
For the outer   
northeastern edge, the temperature rises from $kT_i = 0.71 \pm
0.02$\,keV inside the edge to $kT_o = 1.00 \pm 0.04$ outside the edge, 
such that the temperature jump between undisturbed gas outside the
cold front and gas just within  
the cold front edge is $T_i/T_o = 0.71 \pm 0.05$. Given the density jump of 
$n_i/n_o = 2.9^{+0.5}_{-0.4}$ from Table \ref{tab:edgefits}, 
we find a pressure ratio   
$p_i/p_o = 2.1^{+0.5}_{-0.4}$ across the edge. If this pressure
ratio is interpreted as evidence for bulk motion of the cold front,
then, following Vikhlinin \etal (2001), we infer that the front is
moving transonically  with velocity $\sim 520^{+90}_{-100}$\kms 
(Mach $1.0 \pm 0.2$) relative to the $1$\,keV gas outside the cold
front. 
Thermal pressures across the inner edge are, within the large
uncertainties, consistent with pressure equilibrium and zero 
relative velocity. Although the uncertainties are larger, the pressure 
ratios derived across the southwestern inner (outer) edges are
remarkably similar to those to the northeast. The pressure ratio
across the inner southwestern edge is consistent with thermal
pressure equilibrium, while the pressure ratio across the outer
southwestern edge is $\sim 2$, suggesting possible transonic motion 
of the gas.

\begin{figure}[t]
\begin{center}
\includegraphics[width=3.0in]{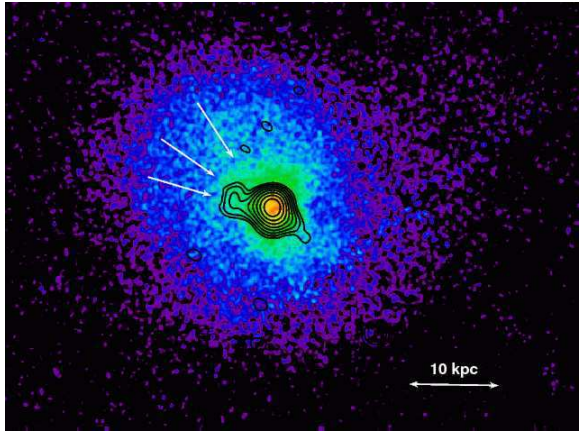}
\caption{{\footnotesize 610 MHz radio contours superposed on the
     background subtracted, exposure corrected, coadded $0.75-2$\,keV 
   Chandra image of diffuse gas in NGC\,5846. Arrows show the location of the
   inner northeastern edge.}
}
\label{fig:610Mhz}
\end{center}
\end{figure}
 
In the upper panel of Figure \ref{fig:edgetprof} we also see that
there is no statistically significant change in the
temperature across the northeastern inner edge. Since these are
projected temperatures, higher temperature gas at large radii along
the line of sight may reduce the observed projected temperature ratio
over that of the true (deprojected) ratio of temperatures across the
front. 

However, the origin of the inner northeastern edge may be more
complex than simple gas sloshing. In the radio band, NGC5846 hosts a 
low radio power ($1.5 \times 10^{21}$\,W\,Hz$^{-1}$ at $1.45$\,GHz)
source (see also Fig. \ref{fig:centralmontage}). Giacintucci et al. (2011) 
observed the source with the Giant Metrewave
Radio Telescope at $610$\,MHz. Their $15^{\prime \prime}$-resolution image
detected a central unresolved component and an extended feature of
only $\sim 3$\,kpc toward the northeast. In Figure \ref{fig:610Mhz} 
we superpose the $610$\,MHz radio emission contours from 
Giacintucci \etal (2011) on the $0.75-2.0$\,kev image of NGC5846, and
find that the inner northeastern edge lies, in projection, just
outside the extended structure detected at $610$\,MHz.   
This suggests the possibility that the inner
northeastern edge may be cool gas displaced by radio plasma in a
bubble formed during a prior episode of AGN activity. 

\begin{figure}[t]
\begin{center}
\includegraphics[width=3.0in]{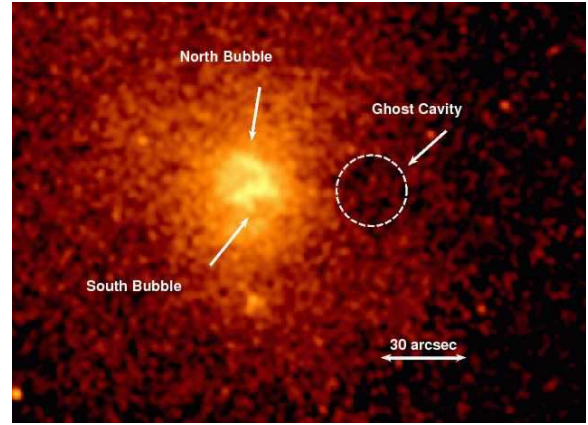}
\caption{{\footnotesize $0.5-2$\,keV close-up view of the central 
$\sim 10$\,kpc}  of NGC\,5846 showing the ghost cavity located $\sim 5.2$\,kpc 
from the nucleus, and the north and south partially disrupted inner 
 X-ray bubbles with centers $\sim 0.6$\,kpc from the nucleus. 
}
\label{fig:centralbubbles}
\end{center}
\end{figure}
\section{X-ray Cavities and Nuclear Activity}
\label{sec:bubbles}

The thermodynamic state of gas in the central $2$\,kpc of NGC\,5846
is complex.  with a non-radial, non-azimuthally symmetric distribution
due both to merging activity and episodic feedback from the central AGN.
Observations in multiple wavebands (radio, optical and X-ray) show
that the gas is truly multi-phase with a mix of cool, dusty filaments,
 warm gas observed through its optical line emission, and hot gas 
with a range of temperatures and densities. The gas is disturbed
both by merging activity and ongoing AGN outbursts, resulting in 
observed non-radial, non-azimuthally-symmetric gas distributions 
that are more difficult to interpret.

\begin{figure*}[t]
\begin{center}
\includegraphics[width=3in]{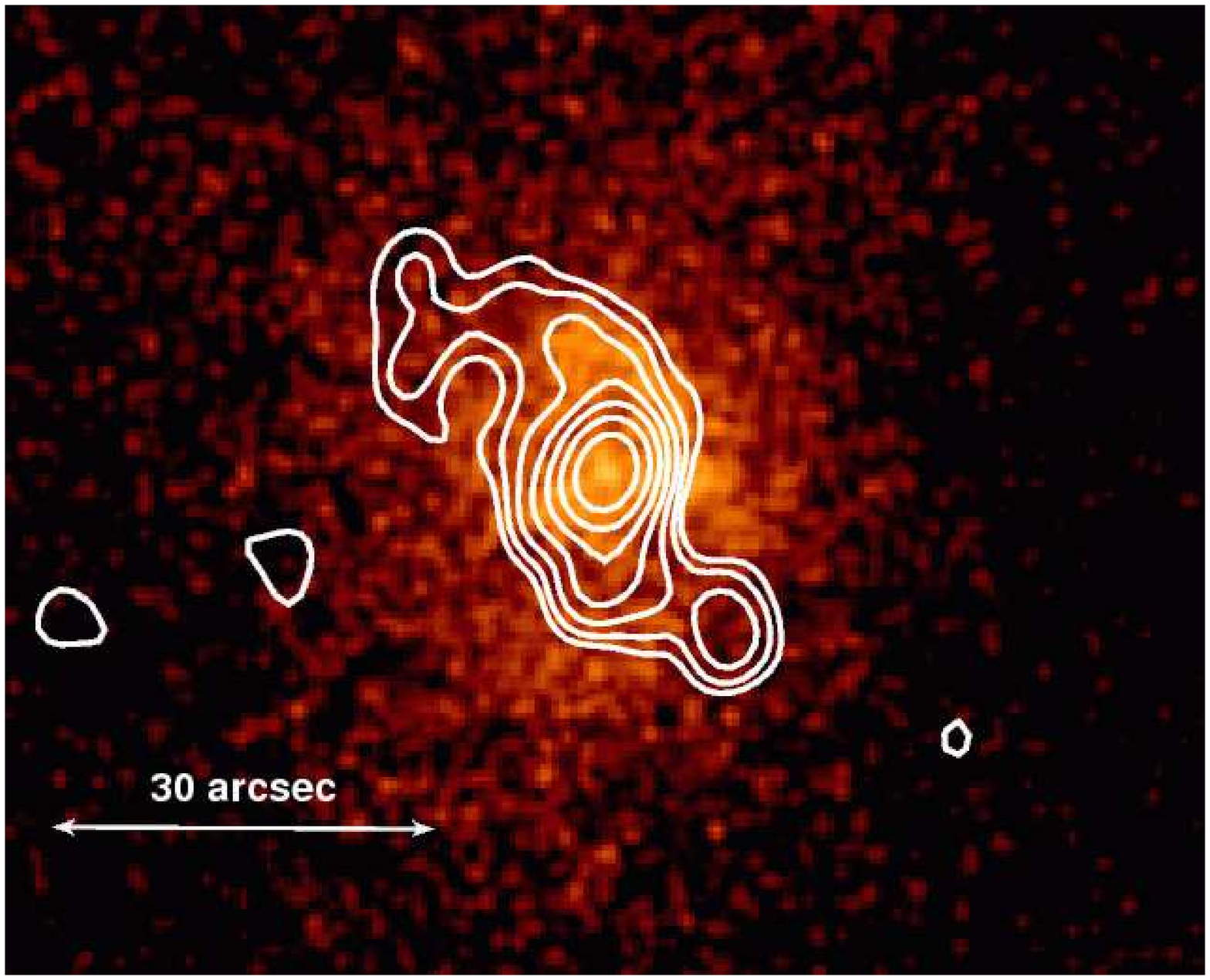}
\includegraphics[width=3in]{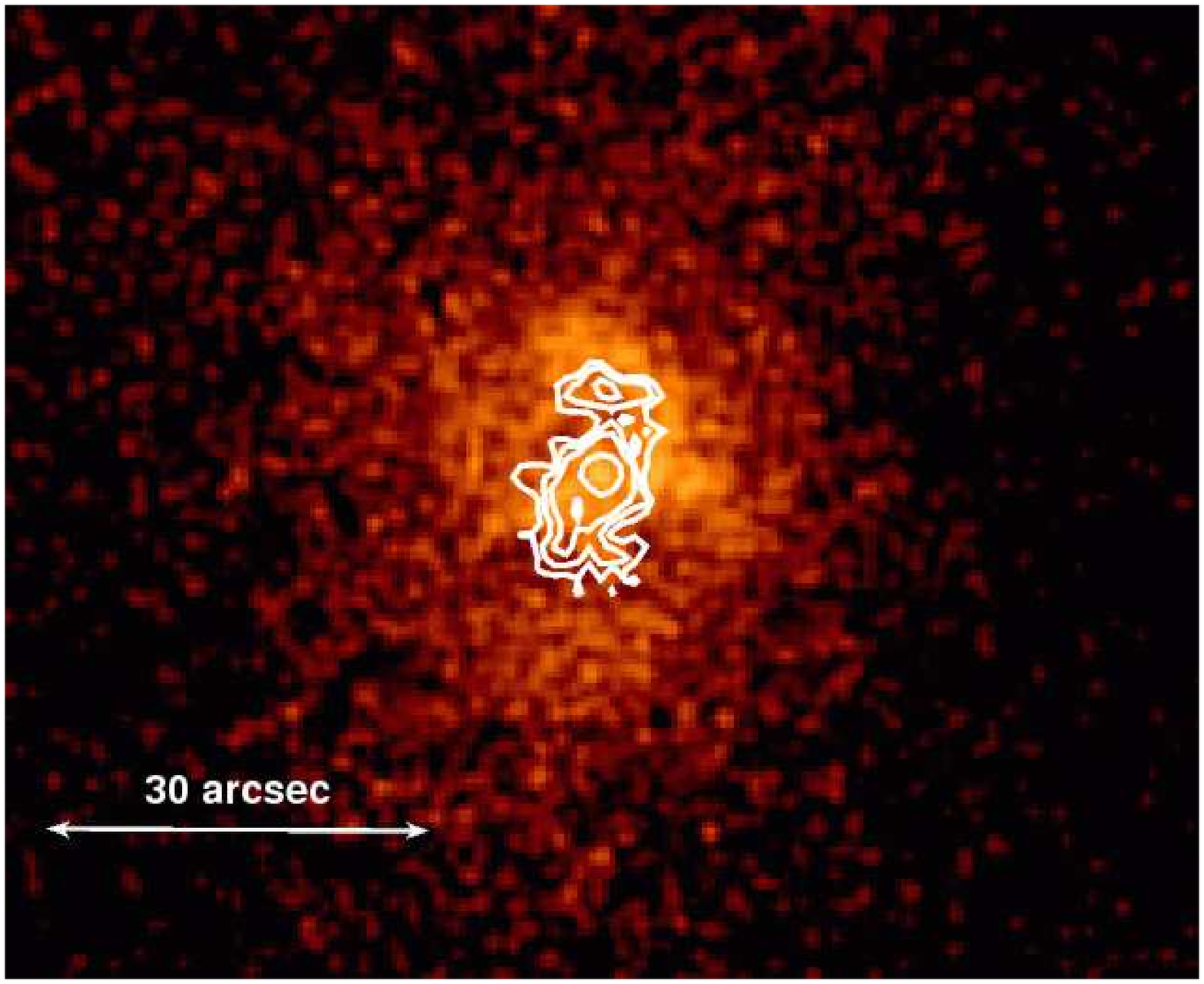}
\includegraphics[width=3in]{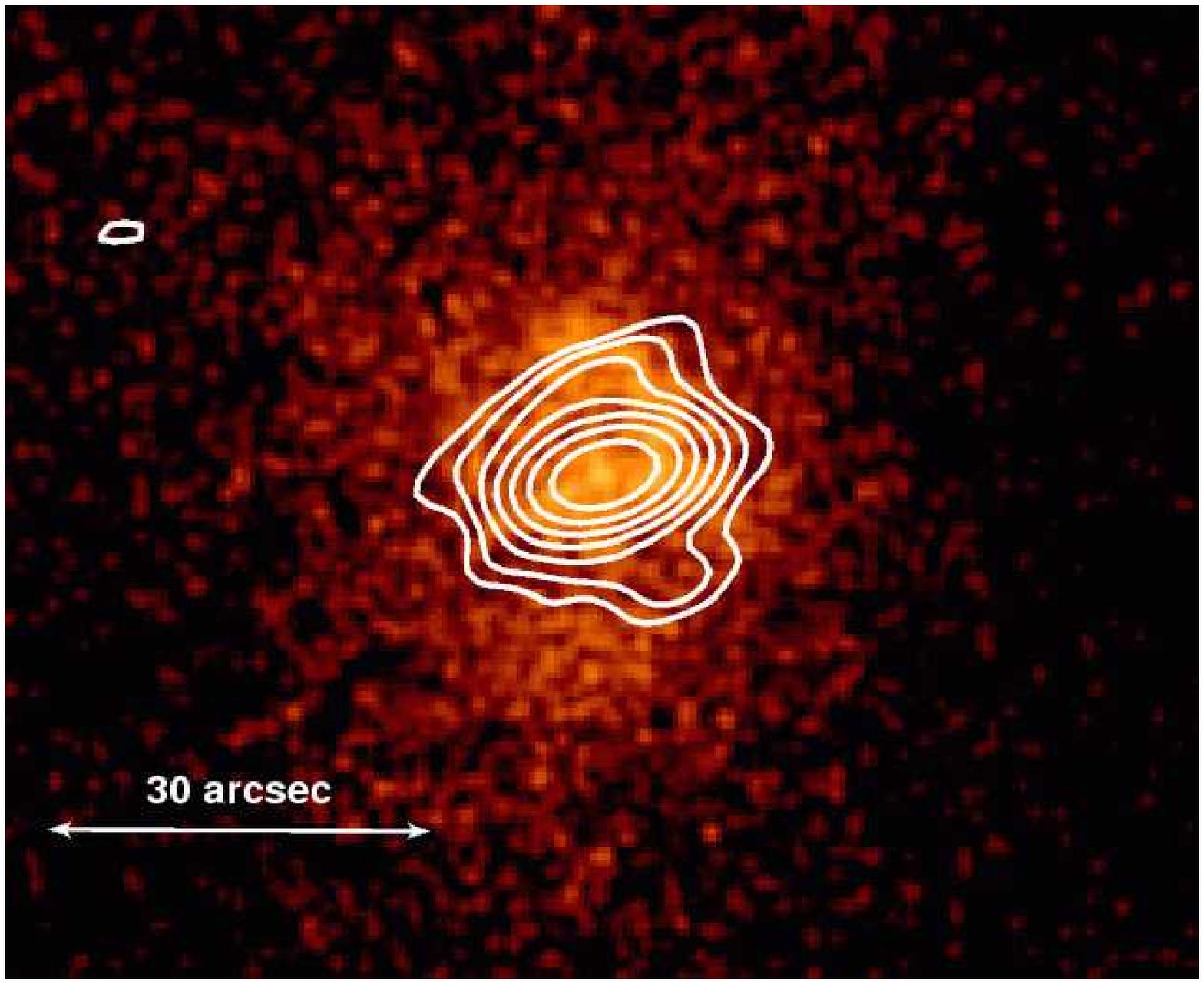}
\includegraphics[width=3in]{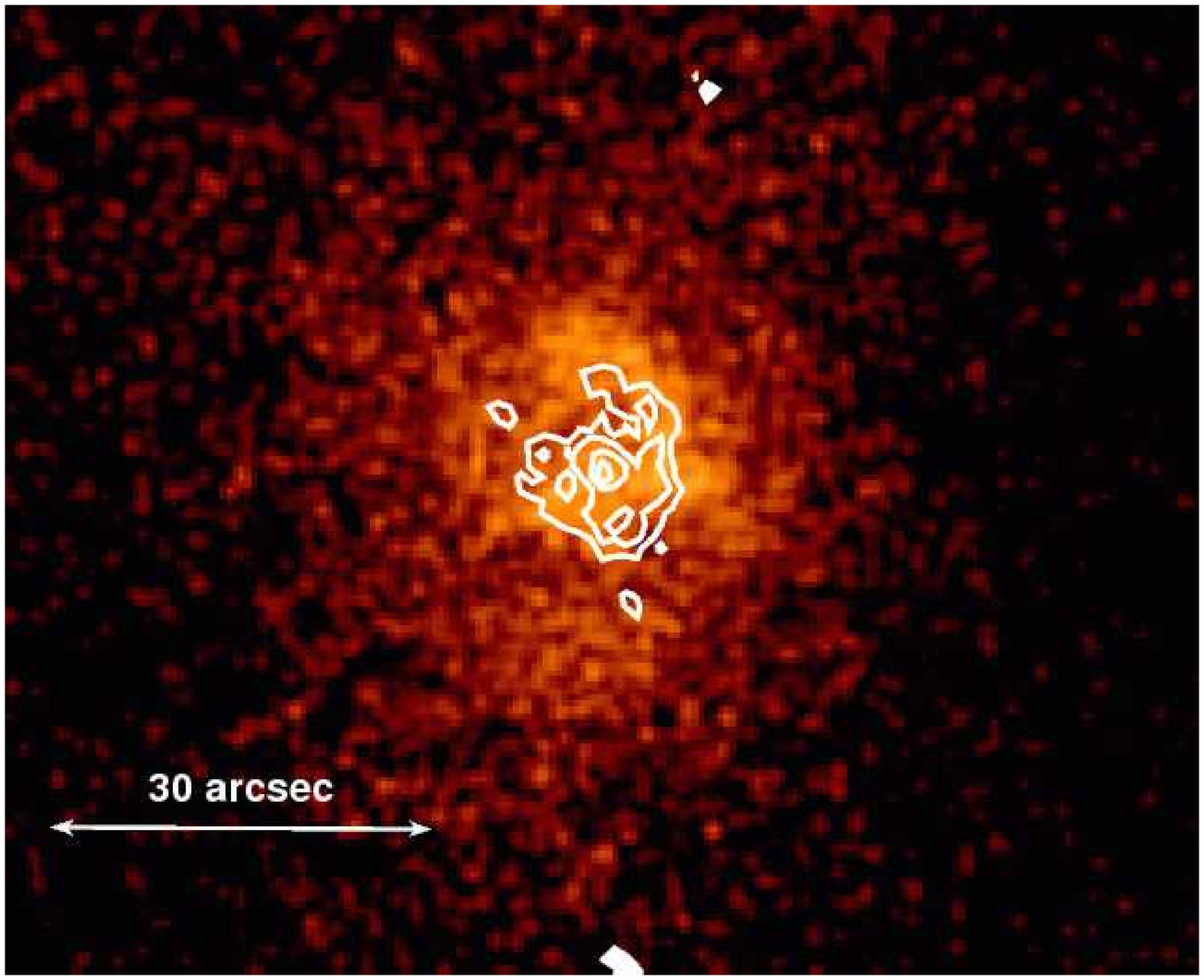}
\caption{{\footnotesize A multiwavelength study of the central
 $4$\,kpc region of NGC\,5846, ({\it upper left}) $0.5-2$\,keV exposure
    corrected, coadded Chandra X-ray image superposed with 
 $1.45$\,GHz VLA radio contours. $1\,{\rm pixel} =
    0\farcs5 \times 0\farcs5$ and the image has been smoothed with
    a $1^{\prime\prime}$ Gaussian kernel. The radio beam has 
    $6^{\prime\prime} \times
    5^{\prime\prime}$ beamsize. Contours begin at $0.11$\,mJy/beam and
    increase by factors of $2$.({\it upper right}) Same Chandra image 
  superposed with higher resolution $1.45$\,GHz VLA radio
    contours to highlight the small scale radio structure. The radio
    beam has $1.5^{\prime\prime} \times 1.5^{\prime\prime}$ beamsize. 
({\it lower left}) The same {\it Chandra} image superposed with
    $5$\,GHz VLA radio contours. The radio beam has $9^{\prime\prime} \times
    5^{\prime\prime}$ beamsize. Contours begin at $0.07$\,mJy/beam and
    increase by factors of $2$.
({\it lower right}) The same {\it Chandra}
    image superposed with H$\alpha$ +[NII]  emission
    contours, from an image taken with the $4.1$\,m Southern Observatory
    Astrophysical Research telescope. 
  Note the close correspondence with small scale structure in the 
  $1.45$\,GHz emission.
}}
\label{fig:centralmontage}
\end{center}
\end{figure*}

Evidence for two recent AGN outbursts are seen in the X-ray `ghost' 
cavity found 
$5.2$\,kpc east of the nucleus, shown
in Figure \ref{fig:n5846full}, and in two roughly symmetrical, 
inner cavities (bubbles), observed previously by 
 Trinchieri \& Goudfrooij (2002) and Allen \etal (2006),   
 whose centers lie $0.6$\,kpc from NGC\,5846's nucleus 
(see Figures \ref{fig:centralbubbles} and 
\ref{fig:centralmontage}).  The inner bubbles  are highly
 significant, with the excess flux in the rims more than $10\sigma$ 
 greater than that expected from background. 
While the rims on the bubbles do not appear
complete (particularly to the south), it is unclear whether this is
the result of actual bubble disruption through the onset of
hydrodynamic instabilities or a projection effect caused by the  
evolution of the bubble through multi-phase gas, already disturbed by 
merging activity. 
 We have retrieved and reanalyzed $1.45$
and $5$\,GHz data from the VLA archive (see Table \ref{tab:radiodat}
for details). Calibration and imaging were carried out using the 
NRAO Astronomical Image Processing System (AIPS) using standard
methods. We find that radio emission at $1.45$\,GHz and $5$\,GHz is 
coincident with the inner bubbles,   
and that the 
$1.45$\,GHz radio 
emitting plasma appears to extend outside the apparently roughly
spherical bubble walls 
to the northeast and south. 
However, no radio emission is detected at $610$\,MHz from the ghost
cavity to the west.

In the upper right panel of Figure \ref{fig:centralmontage}, we show 
contours from an H$\alpha$+[NII] image taken with the $4.1$\,m 
Southern Observatory for Astrophysical Research telescope. 
H$\alpha$+[NII] emission is weak, 
with a total H$\alpha$ flux of $\sim 6.6 \times 10^{-14}$\ergscm. We 
assumed the following ratios between the [NII] lines and the H$\alpha$ line:
${\rm[ NII]}6583/{\rm H}\alpha = 1$ and 
${\rm [NII]}6548/{\rm [NII]}6583 = 0.35$.
With the Kennicutt (1998) relation, where the star formation rate is given
by 
\begin{equation} 
 {\rm SFR} (\Msyr) = 7.9 \times 10^{-42} (L_{{\rm H}\alpha})\,{\rm
 ergs\,s}^{-1},
\end{equation}
the star formation rate in NGC\,5846 is only 
$\sim 0.037\,\Msyr$. The H$\alpha$+[NII] filaments seem dusty and are 
aligned with the X-ray substructures. H$\alpha$ emission is found  
inside the inner edge of the northern bubble rim, and also 
just inside the visible rim of the bubble to the south.  

\subsection{AGN Outburst Energetics} 
\label{sec:ghost}

We estimate the energy of the outburst by the enthalpy ($4pV$) carried 
in the cavity, assuming that the cavity is filled with relativistic 
($\gamma= 4/3$) plasma. $p$ is the mean thermal pressure in the 
undisturbed ambient gas at the same radius as the center of the
bubble, and $V$ is the 
bubble volume (Birzan \etal 2004, Allen \etal 2006). 
The undisturbed gas pressure is estimated using
the mean $\beta$-model fit to the density from \S\ref{sec:slosh} and
temperatures of $0.65^{+0.01}_{-0.02}$\,keV and
 $0.78^{+0.02}_{-0.02}$ for the inner and ghost bubbles,
respectively, taken from the temperature maps at the bubble
locations. For simplicity we assume each bubble is spherical. 
The method used to estimate the bubble age depends on the evolutionary 
stage of the bubble. For young bubbles that are still momentum driven by the
radio jets, the age of the bubble is taken to be the time 
$t_s = d/c_s$ for the bubble to rise to its given location $d$, where
$d$ is the projected distance of the center of the bubble from the
galaxy nucleus, at the speed of sound $c_s$ in the ambient gas 
(Omma \etal 2004; Birzan \etal 2004). For more evolved bubbles, the 
age of the bubble may be better approximated using the time ($t_b$) 
for the bubble to rise buoyantly at its terminal velocity,  
$t_b = d(SC/2gV)^{1/2}$, where $S$ is the
bubble cross-section, $V$ is the bubble volume, $C = 0.75$ is the drag
coefficient, and $g$ is the acceleration of gravity at the projected
distance $d$ of the bubble center from NGC\,5846's nucleus 
(Churazov \etal 2001; Birzan \etal 2004). Once the bubble enthalpy and 
age are determined, the instantaneous mechanical power of the outburst is 
given by  $L_{\rm mech} = E/t$. Our results are given in  
Table \ref{tab:bubbles}, where we have used the buoyancy timescale
$t_b$ to estimate $L_{\rm mech}$. 

For the inner bubbles, we find outburst energies $\sim 10^{55}$\,ergs, 
bubble ages $\sim 2$\,Myr and mechanical powers  
$\sim 2-4 \times 10^{41}$\ergs, consistent with previous work (Allen
\etal 2006). The outburst energy inferred for the `ghost' bubble is
modestly larger ($\sim 5 \times 10^{55}$\,ergs) and the bubble age 
is $\sim 12$\,Myr, giving the time between outbursts (AGN duty cycle) 
$\sim 10$\,Myr. However, the uncertainties in these measurements are
large. First the three dimensional cavity geometry is not known and 
may affect estimates of the bubble volume and, consequently, 
outburst energy by factors of a few. Second, our
simplified model assumes AGN bubbles rising in a static
atmosphere. However, both the inner and ghost bubble ages are shorter
than the expected gas sloshing timescales given by the 
time  since closest approach of the proposed galaxy perturber
NGC\,5850 ($60 - 200$\,Myr; Higdon \etal 1998). 
Thus non-hydrostatic bulk gas motions may have influenced AGN activity and 
likely also affect bubble morphology, the observed distance of the 
bubble from the nucleus, and the probability for bubble disruption. 
Numerical simulations are required to model bubble evolution in this complex,
dynamical environment.
 
\subsection{Knots} 

\begin{figure}[t]
\begin{center}
\includegraphics[width=3.0in]{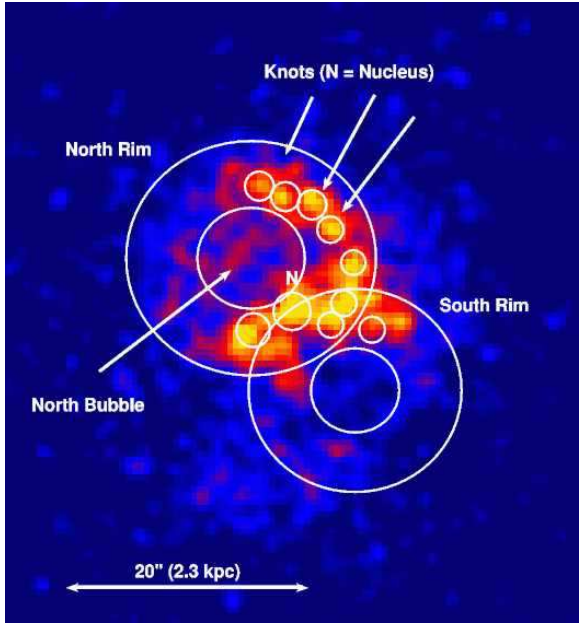}
\caption{{\footnotesize Unsharp masked $0.5-2$\,keV
    background-subtracted, exposure-corrected, coadded Chandra image
    of the inner bubbles in NGC\,5846. The unsharp masking used
    Gaussian smoothing scales of $1$ and $45$\,pixels with an exposure
    map threshold of $20\%$. $1\,{\rm pixel} = 0\farcs5 \times 0\farcs5$.
}}
\label{fig:unsharpknots}
\end{center}
\end{figure}

In Figure \ref{fig:unsharpknots} we use unsharp masking to reveal the
internal structure of the 
northern and southern inner bubbles in NGC\,5846. 
The unsharp masked X-ray image is created using Gaussian smoothing
scales of $1$ and $45$ pixels, respectively,  on the full resolution 
exposure corrected, co-added $0.5-2$\,keV Chandra image. An exposure map 
threshold of $0.20$ was imposed to maximize signal-to-noise in our result.
We see that the inner bubble rims are not smooth, but are 
threaded by $9$ knot-like features (see Table \ref{tab:centralregs}). 
 To determine the detection 
 significance of these knots,  we calculate the 
 cumulative probability that the observed counts or higher in each knot
 in the $0.5-2$\,keV energy band could have arisen from a Poisson 
 fluctuation of the surrounding rim emission.  Knots $4$ and $8$ are 
 least significant with a cumulative probability of $0.02$, while for 
 all other knot features, the cumulative probability that the observed 
 counts or higher  might be a fluctuation of the surrounding rim emission is 
 $< 1.7 \times 10^{-4}$. Thus the significance of the knots as distinct 
 features is high.  

We want to compare the spectra of these knot-like structures to
the spectrum of diffuse gas in the surrounding rims. 
In Table \ref{tab:centralregs} we define circular spectral regions
around each knot and elliptical annuli excluding the knots and nucleus
that trace the north and south bubble rims. An elliptical annulus 
(excluding point sources) located just outside the rims is taken as
the local background for both the knot and rim regions. 
We model the spectrum of the rims with an absorbed VAPEC model
assuming Galactic absorption and fixing the abundances as in 
\S\ref{sec:abundances}. 
As shown in Table \ref{tab:centralspec}, the
mean temperature of gas in the north rim is $0.72^{+0.03}_{-0.02}$\,keV 
for the $89.5$\,ks exposure (OBSID 7923). However, the model does not
provide a statistically acceptable fit ($\chidof =\,111/65$, ${\rm
  null}\,=\,0.03$), and the residuals display the characteristic see-saw
  pattern of multi-temperature gas (Buote 2000). 
In contrast, hot gas in the south rim region is well
modeled by an  absorbed single temperature VAPEC model. 
 We find a  best fit gas temperature of 
$0.64^{+0.02}_{-0.03}$\,keV 
($\chidof=\,94.3/97$), where we have used a simultaneous fit to both the
$89.5$\,ks (OBSID 7923) and $23.2$\,ks (OBSID 788) data sets to reduce 
uncertainties. The more complex spectral structure for gas in the north rim
compared to that in the south rim region may be due to projection
effects, such that multi-phase gas is more prominent along our line of
sight to the north than to the south. Alternatively, the combination of AGN and
merging activity may have introduced hydrodynamic instabilities in the
fluid that were more efficient at disrupting the bubble to the south
than to the north.  The intrinsic $0.5-2$\,keV luminosity 
of diffuse gas in the north rim ($1.0 \times 10^{40}$\ergs) is $50\%$
higher than that in the south rim ($0.65 \times 10^{40}$\ergs).

The individual knot-like features are X-ray faint.  
In Table \ref{tab:netcounts} we list the net source counts for the
individual knots in four energy bands: $0.6-0.8$\,keV (FeSS),
$0.8-1.1$\,keV (FeS), $1.1 - 2.0$\,keV (M), $2-4$\,keV (H). 
The first two bands are chosen to divide the energies containing 
the FeL line complex 
(the Fe peak from $0.6 - 1.1$\,keV) in half, such that the hardness ratio 
constructed from these bands will be sensitive to shifts in the energy of the 
peak emission from the FeL complex, caused by temperature differences in 
cool ($\lesssim 1$\,keV) gas.  
For comparison, we also include net source counts in these energy bands for
the north and south rims. Notice that in all cases (knots and rims), 
the X-ray emission is dominated by the Fe peak, confirming that the
X-ray emitting gas is cool, and, for all the knot-like features  
except Knot$9$, the net photon counts above $2$\,keV 
are consistent with zero at the $\lesssim 2\sigma$ confidence
level. Since the number of photons in each individual knot is small, 
we use the  hardness ratio across the iron peak, 
$({\rm FeS} - {\rm FeSS})/({\rm FeS} + {\rm FeSS})$, 
 to search for  temperature differences between the individual knots and
 between the knots and the rims. We also compute the standard medium
 band hardness ratio $({\rm M} -{\rm  S})/({\rm S}+{\rm M} +{\rm H})$, 
where S is the number of photons in the $0.6 - 1.1$\,keV energy
band. These hardness ratios along with the total ($0.6-4$\,keV) net 
source counts for each region are given in Table \ref{tab:hardratio}. 
Although the hardness ratios for Knot$4$ and Knot$5$ suggest a mildly
softer spectrum than the others, the most striking result from 
Table \ref{tab:hardratio} is that the hardness ratios of the knots are
nearly identical to that of the rims. We stack the knot spectra to determine 
quantitatively a mean temperature for the knots, and confirm that the
best fit temperature ($0.72^{+0.4}_{-0.7}$, $\chidof=\,67.3/45$ for
the $89.5$\,ks exposure) is statistically indistinguishable from that 
measured in the surrounding rims.
The total $0.5-2$\,keV luminosity of the knots is 
$0.3 \times 10^{40}$\ergs, one third of the X-ray luminosity of
diffuse gas in the north rim. 

To investigate the gas density and thermal pressure in the knots, we
fit the spectrum of the brightest knot (Knot$6$ in the north rim 
with $0.5-2$\,keV luminosity of $5.7 \times 10^{38}$\ergs) as an example. 
We fix the VAPEC model temperature to $0.72$\,keV, 
the best fit temperature for the stacked knot spectrum (and 
gas in the north rim), and allow only the VAPEC model normalization 
to vary. We then compute the electron density from the VAPEC
normalization (see eq. \ref{eq:napec}), assuming that the knot is a 
sphere with radius  $1\farcs4$ ($0.162$\,kpc) uniformly filled with
gas. We find an electron density of $0.33 \pm 0.03$\cmc and
corresponding thermal pressure of $7.3 \times 10^{-10}$\ergscmc where the 
uncertainties reflect only the statistical uncertainties of the
spectral model fit. For such dense gas, the cooling time is short, 
$\sim 30$\,Myr.

For comparison, we compute the electron density and corresponding 
thermal pressure for the diffuse gas in the north rim 
surrounding Knot$6$. Again the largest
uncertainty in the analysis is the assumed morphology of the emission
region. To bound the electron density in the rim, we compute the
electron density assuming first that the observed emission is gas
uniformly distributed throughout a cylindrical shell with a cross 
sectional area that of the spectral extraction region
and a height equal to the bubble radius (an upper limit on the emission
volume and consequently lower limit on the derived density) and,
second, that the X-ray emission comes solely from a spherical shell 
with dimensions determined by the spectral extraction region (a lower
limit on the emission volume), excluding the knot regions in each
case. The electron density in the north rim is then bounded by 
$0.076 \leq n_e \leq 0.096$\cmc. This is somewhat larger, as
expected by the greater surface brightness of the rim, but comparable to the 
electron gas density ($0.074$\cmc) derived from the mean $\beta$-model 
for NGC\,5846 (see \S\ref{sec:slosh}). We infer a thermal pressure in the
gas surrounding Knot$6$ of $ (1.7 -2) \times 10^{-10}$\ergscmc.
at least a factor $3$ lower than that derived for Knot$6$, assuming
a spherical geometry. 

There are two possible scenarios to explain this discrepancy. 
If the knots are stable structures in pressure equilibrium with their
surroundings, then either the knots are not spherical,
cooling blobs or they are stabilized by non-thermal pressures, or
both. In this scenario, the most likely explanation is that the 
true $3$-dimensional morphology of the knot-like structures is more 
extended and complex. For example if they are filamentary rather than 
spherical, i.e. more like the X-ray filament observed in M87 
(Forman \etal 2007) but
viewed with the filament oriented along the line of sight, 
the emission volume would be larger, implying a lower electron density
and allowing the structure to be closer to or in thermal equilibrium with
its surroundings. Then the  knot-like appearance of the structures would
be primarily a projection effect. It may be  difficult to explain,
however, why all the knots would have this orientation.

The second possibility is that the pressure estimates are, in fact,
roughly correct and the knots and surrounding features are the visible
result of a strong shock being driven into the ambient gas by the
supersonic inflation of the radio lobe.  The general morphology of the
rim and knots is similar to the high Mach number shocks seen in NGC 4552
(Machacek et al 2006) and Her A (Nulsen et al. 2005).  There is a
noticeably bright, roughly elliptical, enhancement associated with the
radio lobes in both cases.  The morphology of the knots and the rim
looks similar, albeit at much lower signal to noise.  The knotty
morphology could be the result of a shock being driven into a
multi-phase ISM.  In this scenario, the knots are the result of cooler
clouds embedded in a hotter medium.  A shock driven through such a
stratified medium will drive spherical shocks into the clouds.  They are
brighter because they are denser, while the post-shock gas must be in rough
pressure equilibrium.  

We can use the density  between the rim and the pre-shocked 
gas in the Rankine-Hugoniot conditions to estimate the Mach number of 
the shock and the expected temperature increase in the rim 
(Landau \& Lifschitz 1959). 
Assuming the bubble rims are spherical 
shells and that the preshock density is given by our mean $\beta$-model from 
\S\ref{sec:slosh} evaluated at the midpoint of the rim, we can use the 
density jump between the rim and the pre-shocked gas in the Rankine-Huygenot 
conditions to estimate a  shock Mach number. From the more complete 
northern rim, we find a density jump of $1.4$, and  Mach number $1.3$. 
However, the Rankine-Huygenot conditions assume the shock front is thin.   
Our measured value for the density jump is an average value  across
the rim, and so  will underestimate the true density jump at the shock 
front. Thus this inferred Mach number should be interpreted as a lower 
bound on the true Mach number for the shock. These averaged values for 
the north rim in NGC\,5846 are similar to those measured  
for bubbles in the central region of NGC\,4552, where more detailed modeling 
indicated a shock Mach number of $\sim 1.7$ (Machacek  \etal 2006). 
The measured (projected) temperature jump across the northern rim 
in NGC\,5846 ($\sim 1.1$) is expected to be smaller than the Rankine-Hugoniot 
prediction ($1.3$) due to dilution by emission from unshocked gas along the 
line of sight. Note that knotty features are also seen in X-ray images of 
 NGC\,4552's shocked bubble rims (see, e.g, the right panel of Fig. 1 in 
Machacek \etal 2006).

\section{NGC5846A: Galaxy Infall \& Gas Stripping}
\label{sec:N5846A}

\begin{figure}[t]
\begin{center}
\includegraphics[width=3.0in]{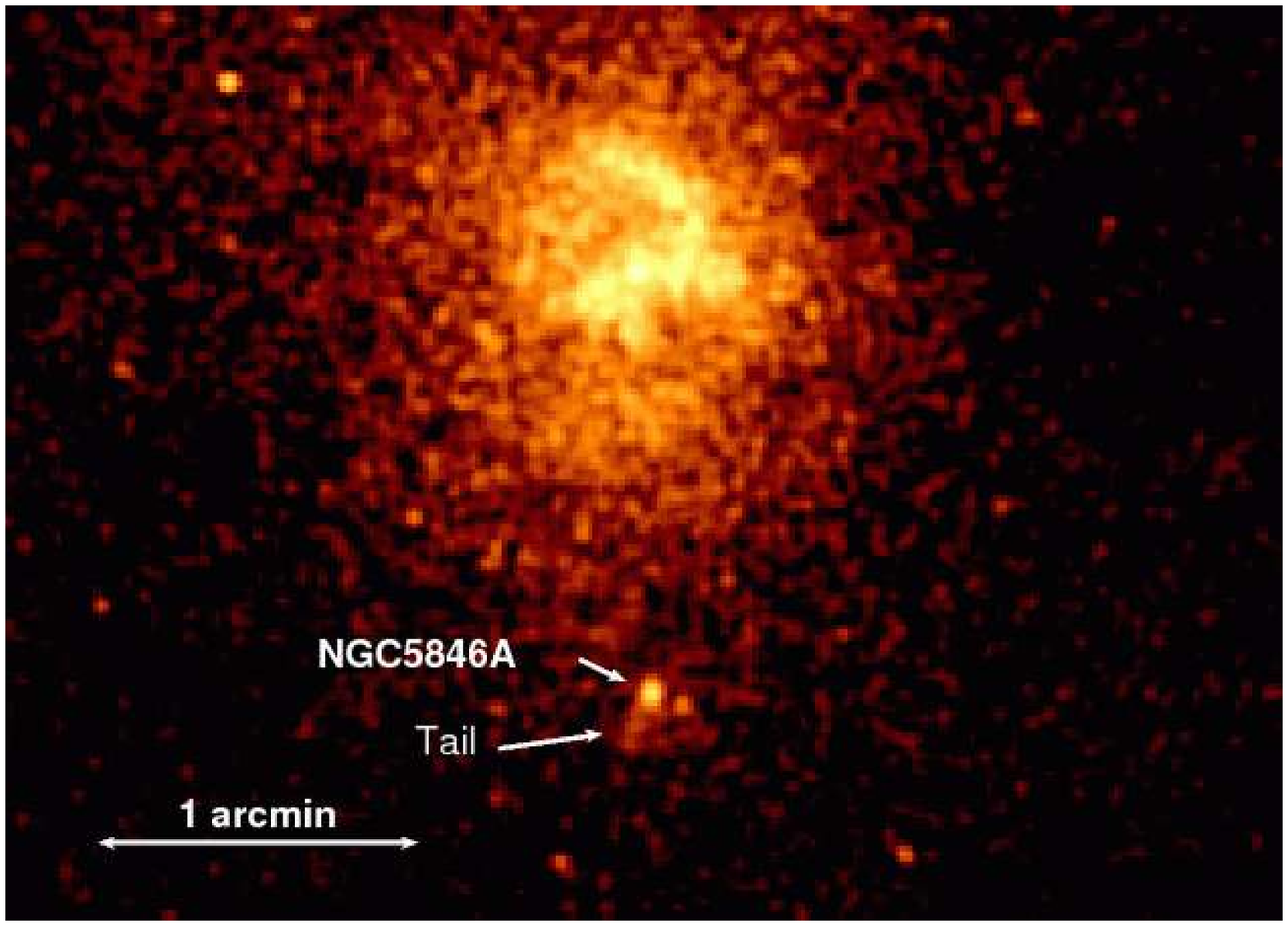}
\includegraphics[width=3.0in]{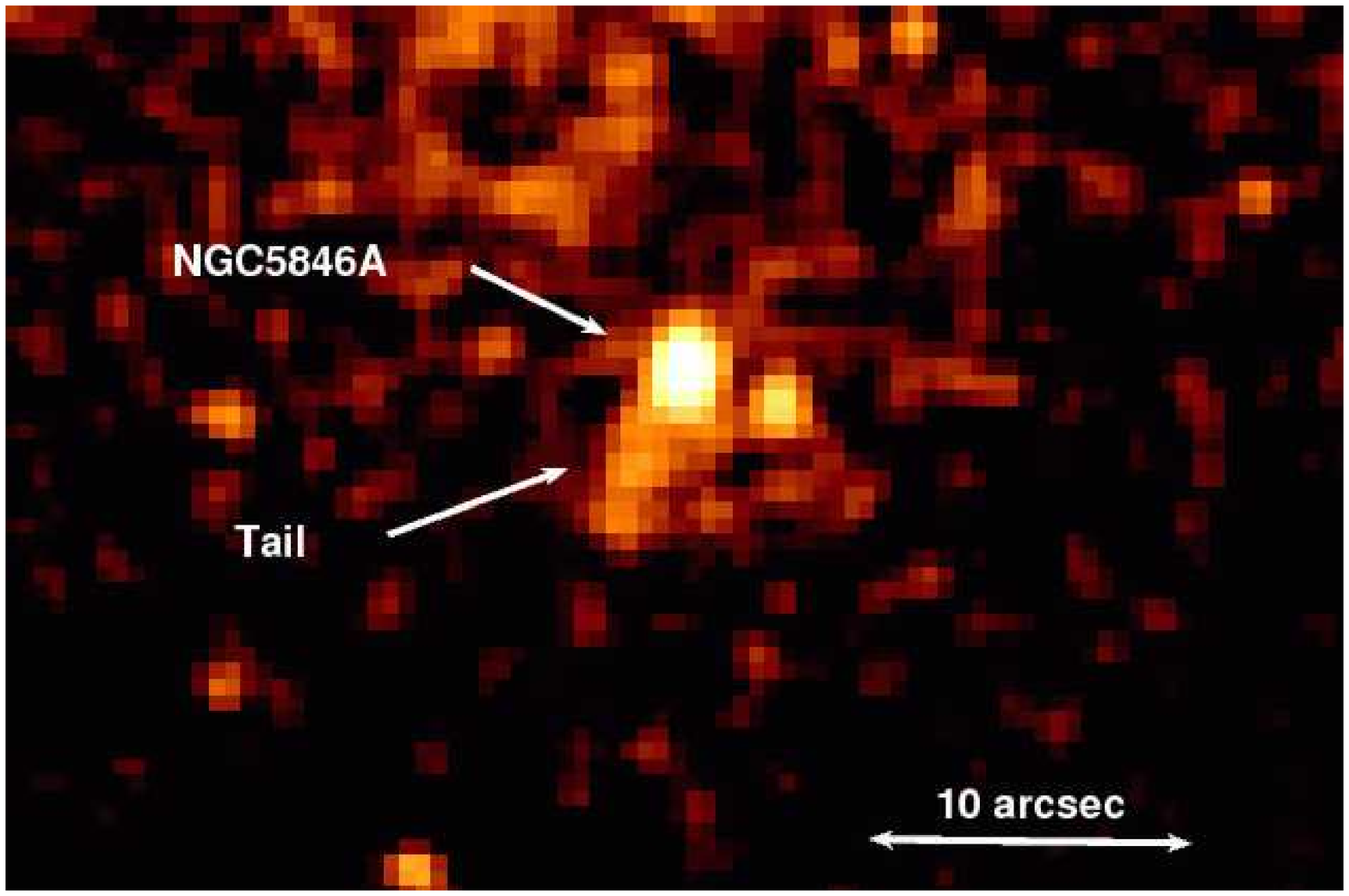}
\caption{{\footnotesize ({\it upper}) $0.3-2$\,keV background-subtracted,
    exposure-corrected, coadded Chandra image of the center of
    NGC\,5846 showing the infalling cE galaxy NGC\,5846A. ({\it lower}) 
     Close-up of the image in the left panel highlighting 
    NGC\,5846A's small gaseous corona and stripped
    tail. $1 pixel = 0\farcs5 \times 0\farcs5$ and the image has
    been smoothed with a $1\farcs0$ Gaussian kernel. }
}
\label{fig:ngc5846A}
\end{center}
\end{figure}

NGC\,5846A, shown in Figure \ref{fig:n5846dss}, is directly 
south of the dominant group elliptical galaxy NGC\,5846.
NGC\,5846A was among the earliest galaxies confirmed to be a compact
elliptical galaxy (cE). Compact elliptical galaxies form a rare class 
of early type galaxies that share 
many of the unusual properties of M32, companion to M31 (Andromeda) 
in the Local Group (Faber 1973). Compact elliptical (cE)  galaxies have low 
optical luminosities ($\sim 10^9\Ls$), similar to the more common dwarf
elliptical galaxies, but their half-light radii ($\sim 0.25$\,kpc) are 
several times smaller, such that their optical surface brightness is
much higher than expected for their luminosity (e.g. Faber 1973; 
Smith Castelli \etal 2008; Chiligarian \etal 2009). All are found near 
massive companions. They also  tend to have higher stellar velocity 
dispersions and higher metallicities than expected for their
luminosities, and lie above the locus of normal dwarf elliptical galaxies on
the Faber-Jackson relation (Faber \& Jackson 1976).
Currently only about a dozen cE galaxies have been
found within $100$\,Mpc, where their properties can be confirmed  by
ground-based telescopes (for recent searches see  
Smith-Castelli \etal 2008, 2010 in the Antlia cluster; 
Price \etal 2009 in Coma; Chiligarian \& Bergond 2010 in the NGC\,5846
group). 

While the origin of cE galaxies and their relationship to galaxy 
evolution remain under debate, the most likely 
explanation for their high surface brightness, high velocity
dispersion and compact structure  is
that their progenitors were  massive, possibly disky, galaxies 
that have undergone severe tidal threshing from interactions with a 
 nearby massive companion. The outer stellar layers have been tidally 
stripped from these galaxies leaving only the compact stellar core
(Bekki \etal 2001, 2003).  The near-infrared indices for NGC5846A  
suggest that the progenitor of NGC5846A may have 
been a massive elliptical galaxy, and that the catastrophic tidal 
interaction, responsible for stripping $\sim 90\%$ of its progenitor's 
stellar mass, may also have funneled gas into NGC5846A's core, 
resulting in an intermediate age population of stars near its nucleus
(Davidge \etal 2008).  

One might expect such strong tidal 
interactions would also remove any hot gas corona associated with 
the compact elliptical galaxy, leaving it devoid of X-ray emitting
gas.  
Figure \ref{fig:ngc5846A} shows, however,   
an X-ray tail extending $0.54$\,kpc ($4\farcs7$) to the south of
NGC5846A, evidence for ongoing ram pressure stripping of a 
residual gaseous corona of the compact elliptical galaxy. 
Since the radial velocity difference between NGC\,5846A and NGC\,5846 
is $\sim 487$\kms (NED), already transonic for $\sim 0.7$\,keV gas, 
the infall is likely supersonic. Assuming a thermal model with $kT
\sim 0.5 -1$\,keV, uniform filling  and cylindrical geometry 
for the tail, the mean density of gas in the tail is estimated to be 
$\sim 0.05$\cmc and mass in the tail of $\sim 10^5\,\Ms$. 
%$n_e = 0.063$\cmc and $1$\,keV gas. 
The average mass return rate for a simple population
of stars in a  galaxy of age $t_*$, where $t_*$ is the end of the 
galaxy formation epoch and stars are passively evolving, is  
$dm_*/dt \sim 1.5 \times 10^{-11}(t_*/15)^{-1.3}\,L_{\rm
  B}\Msyr$ (Ciotti \etal 1991). Assuming that NGC5846A shares many of
the same properties as the compact elliptical galaxy NGC4486B 
(Davidge \etal 2008), such
that $t_* \sim 12$\,Gyr (Soria \etal 2006), and has an  
absolute B-band magnitude $\sim -17$ (Davidge 1991), the rate of mass
loss by aging stars is $\sim 0.018\Msyr$. The time between stripping
episodes  needed to replenish the ISM with sufficient mass to explain
the gas mass observed in the tail is  $\sim 5.5$\,Myr.
 
The central X-ray corona in NGC\,5846A has a radius $\sim 1\farcs3$, 
also of the order of the Chandra point spread function. Since this 
is  consistent with a point source, we can only set an
upper limit of $\lesssim 150$\,pc for the physical radius of
NGC\,5846A's gas corona. Interpretation of the X-ray properties 
of the central region of NGC\,5846A is further complicated by the 
presence of a nuclear point source. The observed $2-10$\,keV source 
count rate in a $1\farcs3$ circular region surrounding the nucleus 
($\sim (3.4 \pm 0.6) \times 10^{-4}$\cnts) is more than an order of 
magnitude larger than that expected if X-ray emission in the region
was due solely to hot the X-ray emission from 
the region were due to $\sim 1$\,keV hot gas with solar abundance. 
Assuming a power law model with photon index $\Gamma = 1.4$ for the 
nonthermal emission, we find a $2-10$\,keV nuclear luminosity of
$6.6\times 10^{38}$\ergs. Extrapolating this power law model for 
the nuclear emission to the
$0.5-2$\,keV band, we find that the observed count rate  
exceeds that predicted by the model by $\lesssim 2.4\,\sigma$. 
If we assume this excess $0.5-2$\,keV emission is due entirely to 
a $0.5-1$\,keV hot gas corona with solar abundance that uniformly 
fills a spherical volume  of radius $150$\,pc, we find an upper limit
on the  mean electron  density and gas mass of the gas corona of 
$\sim 0.1$\cmc and  $\lesssim 3 \times 10^4\,\Ms$, respectively. 
These should be considered upper limits,
because if the nuclear spectrum is steeper, the nuclear contribution 
to the $0.5-2$\,keV emission increases.  For $\Gamma \gtrsim 1.7$, 
the nuclear point source alone is sufficient to explain all of the 
observed soft emission from the central region. Thus most of the hot 
gas associated with NGC5846A resides in the stripped tail.

\section{CONCLUSIONS}
\label{sec:conclude}

NGC\,5846 may be the best nearby example of a galaxy group in which we 
can simultaneously study the dynamics of non-hydrostatic gas motions
induced by galaxy interactions, AGN activity and bubble evolution. 
In this paper we have used the combined $120$\,ks {\it Chandra}
exposure for NGC\,5846 to analyze the signature X-ray surface
brightness edges and cavities produced by this activity, and have
compared these data with radio and H$\alpha$ images of the same
region. We also observed for the first time ram pressure stripping 
of a  compact elliptical galaxy NGC\,5846A, during its supersonic 
infall towards NGC\,5846. We find: 

\begin{itemize}

\item{The surface brightness distribution for NGC\,5846 is
  asymmetrically displaced northeast from the nucleus, 
  displaying surface brightness discontinuities (edges) 
  at $19.8$\,kpc and $6.7$\,kpc from the nucleus to the northeast 
  and at $19.1$\,kpc and $11$\,kpc to the southwest. Assuming a double
  power law model for the electron density and no strong abundance
  gradients, the density jump at the outer (inner) edges 
  are $2.9^{+0.5}_{-0.4}$ ($1.4 \pm 0.2$) for the edges to the
  northeast and $2.1^{+0.7}_{-0.4}$ ($1.6 \pm 0.2$) to the
  southeast. 
 These multiple surface brightness
  edges are likely the signature of non-hydrostatic gas
  motions (sloshing) caused by the off-axis encounter of NGC\,5846 
  with another group  
  galaxy, possibly the large spiral NGC\,5850, or a merging subgroup.}  

\item {Temperature maps for NGC\,5846 confirm that the observed
  surface brightness edges  correspond to cold fronts. Spectral
  modeling with a VAPEC thermal plasma model across the northeast and
  southwest outer edges give temperature ratios of $\sim 0.7$ and
  $0.9$, respectively, and pressure ratios $\sim 2$ suggesting that
  the gas in the cold front may be moving transonically relative to
  its surroundings.  Thermal pressures across the inner
  edges are consistent with pressure equilibrium and zero relative
  velocity.}

\item{ The surface brightness map of NGC\,5846 shows a
  spiral-like tail extending $38.5$\,kpc to the northwest. Such 
  spiral structures are also characteristic of gas sloshing. However,
  no corresponding spiral structure is observed in the temperature
  map. This suggests that the perturbing encounter did not occur in
  the plane of the sky, and may be consistent with the suggestion by
  Higdon \etal (1998) that NGC\,5846 recently experienced a close
  off-axis interaction with the spiral galaxy NGC\,5850.}

\item{Evidence for recent AGN activity in NGC\,5846 are seen in two
 roughly symmetrical bubbles with centers $\sim
 0.6$\,kpc from the nucleus,  and in a ghost X-ray cavity $5.2$\,kpc to
 the east.  The inner bubbles are coincident with $1.45$\,GHz and
 $5$\,GHz radio emission. 
 H$\alpha$ emission along the inner edge of the northern bubble
 rim suggests that gas is cooling, supporting the complex, multi-phase nature
 of gas in the central region.  However,
 the H$\alpha$ emission is weak ($\sim 6.6 \times 10^{-14}$\ergscm), such that 
 the inferred star formation rate is only $\sim 0.037\,\Msyr$.}

\item{From the bubble morphology and thermodynamic properties of the 
surrounding gas, we infer outburst energies, bubble age, and
AGN mechanical luminosity of $\sim 10^{55}$\,ergs, $\sim 2$\,Myr, and 
$(2-4) \times 10^{41}$\ergs, respectively, for the inner bubbles, 
consistent with previous results. The outburst energy and age of 
the ghost cavity are  $\sim 5 \times 10^{55}$\,ergs and $12$\,Myr, 
implying an AGN duty cycle of $\sim 10$\,Myr. Both the inner bubble 
and ghost cavity ages are shorter than the time since NGC\,5850's 
closest approach to NGC\,5846, such that non-hydrostatic
bulk gas motions from the encounter may have influenced AGN activity, 
bubble morphology and position, and possibly bubble disruption 
in NGC\,5846.} 

\item{The inner bubble rims are threaded with $\sim 9$ 
bright, knot-like features with individual $0.5-2$\,keV luminosities 
$\lesssim 6 \times 10^{38}$\ergs. The total  $0.5-2$\,keV 
X-ray luminosity of these `knots' is $0.3 \times 10^{40}$\ergs, 
compared to X-ray luminosities of $1.0 \times 10^{40}$\ergs and $0.65
\times 10^{40}$\ergs for diffuse gas in the north and south rims,
respectively. The mean temperature derived from modeling the stacked 
spectrum of the `knots' ($0.72^{+0.4}_{-0.7}$), is statistically
indistinguishable from that of the surrounding gas in the
rim. The electron density and thermal pressure in the 
brightest `knot', derived assuming the 'knot' is spherical, is at least 
a factor three higher than that in the surrounding rim gas. We suggest 
either the `knots' are extended, filamentary structures in equilibrium 
with the surrounding gas or they are transient structures, i.e. gas
clouds heated and compressed by the recent passage of a shock from the last 
AGN outburst.   
}      

\item{{\it Chandra} images of the compact elliptical galaxy NGC\,5846A, 
infalling supersonically towards NGC\,5846 from the south, show a
compact ($\lesssim 150$\,pc radius) corona at the galaxy center and a 
$0.5$\,kpc X-ray tail containing $\sim 10^{5}\,\Ms$ of gas,
suggesting ongoing ram-pressure stripping. Assuming that the observed
gas originated from passively evolving stars, the time between
stripping episodes needed to replenish the galaxy ISM with sufficient
mass to explain the gas mass in the tail is $\sim 5.5$\,Myr.}

\end{itemize}

%----------------------------------------------------------------------------

\acknowledgements

This work is supported in part by NASA grant NNX07AH65G
and the Smithsonian Institution. 
This work has made use of the NASA/IPAC Extragalactic Database (NED)
which is operated by the Jet Propulsion Laboratory, California
Institute of Technology,  under contract with the National
Aeronautics and Space Administration. We would like to thank John
ZuHone and Ryan Johnson for helpful discussions. 

\noindent{\it Facilities:} CXO (ACIS-I, ACIS-S), VLA
%--------------------------------------------------------------------------

\begin{small}

\end{small}

\begin{deluxetable}{ccccc}
\tablewidth{0pc}
\tablecaption{Azimuthally Averaged Spectral Fits\label{tab:tprof}}
\tablehead{\colhead {$r_i$,$r_o$} & \colhead{$r_{\rm mean}$} & \colhead{$kT$} 
  & \colhead{$\chi^2/{\rm dof}$}  & \colhead{null}\\
 (arcsec, arcsec) & (kpc)  &(keV) &  & }
\startdata
  $0$,     $6$   &$0.35$  &$0.72^{+0.02}_{-0.03}$  & $76/48$ &$0.006$\\
  $6$,    $13.2$ &$1.13$  &$0.66^{+0.02}_{-0.02}$  & $143/66$& $10^{-7}$\\
 $13.2$,  $21.8$ &$2.03$  &$0.64^{+0.02}_{-0.02}$  & $72/66$ & $0.29$\\
 $21.8$,  $32.2$ &$3.13$  &$0.62^{+0.02}_{-0.02}$  & $64/67$ &$0.58$\\
 $32.2$,  $44.6$ &$4.45$  &$0.63^{+0.01}_{-0.02}$  & $56/68$ & $0.86$\\
 $44.6$,  $59.6$ &$6.04$  &$0.65^{+0.02}_{-0.02}$  & $92/72$ &$0.06$\\
 $59.6$,  $77.5$ &$7.95$  &$0.70^{+0.01}_{-0.01}$  & $105/74$& $0.01$ \\
 $77.5$,  $99.0$ &$10.24$ &$0.73^{+0.01}_{-0.01}$  & $89/78$ &$0.18$\\
 $99.0$, $124.8$ &$12.98$ &$0.75^{+0.01}_{-0.01}$  & $81.7/82$& $0.49$\\
$124.8$, $155.8$ &$16.27$ &$0.83^{+0.02}_{-0.02}$  & $110/82$& $0.02$\\
$155.8$, $192.9$ &$20.22$ &$1.01^{+0.02}_{-0.03}$  & $100/84$&$0.11$ \\
$192.9$, $237.5$ &$24.96$ &$1.15^{+0.06}_{-0.11}$  & $113/85$& $0.02$\\
$237.5$, $291.0$ &$30.65$ &$1.24^{+0.07}_{-0.08}$  & $74.6/95$&$0.94$ \\
$291.0$, $355.2$ &$37.48$ &$1.23^{+0.09}_{-0.11}$  & $89/99$&$0.75$ \\
\enddata
\tablecomments{\footnotesize{Azimuthally average spectra for the
NGC\,5846 group from {\it Chandra} OBSID 7923,  taken in circular 
annuli centered on the nucleus of NGC\,5846 and with logarithmically 
increasing bin width. $r_i$, $r_o$ and $r_{\rm mean}$ are the inner,
outer, and mean radius of each annulus. All spectra,except the central
bin, are modeled with an absorbed VAPEC model with fixed Galactic absorption 
($4.24 \times 10^{20}$\cmc), and abundances for O, Mg, Si,and Fe 
of $0.24$, $0.53$, $0.72$, and $0.49\,\Zs$, respectively.  
All other abundances are fixed at $0.5\,\Zs$ (see 
\S\protect\ref{sec:abundances}). }
}
\end{deluxetable}

\begin{deluxetable}{cccccccc}
\tablewidth{0pc}
\tablecaption{Angular Sector Density Model Fits\label{tab:edgefits}}
\tablehead{\colhead{Sector} & \colhead{$r_1$} &  \colhead{$J_1$} 
& \colhead{$r_2$} & \colhead{$J_2$} & \colhead{ $\alpha_1$} 
  & \colhead{$\alpha_{{\rm mid}}$}& \colhead{ $\alpha_2$} \\
 & (kpc) & & (kpc) & & & & }
\startdata
NE & $6.6-6.8$ & $1.4^{+0.2}_{-0.2}$ & $19.8$ 
 & $2.9^{+0.5}_{-0.4}$ & $-0.5^{+0.4}_{-0.4}$ & $-0.61^{+0.07}_{-0.07}$ 
 & $-1.2^{+0.1}_{-0.2}$ \\
SW & $11.0$ & $1.6^{+0.2}_{-0.3}$ & $19.1$ & $2.1^{+0.7}_{-0.4}$ 
& $-0.7^{+0.2}_{-0.4}$ & $-0.8^{+0.5}_{-0.4}$ &
   $-1.2^{+0.2}_{-0.2}$ \\  
\enddata
\tablecomments{\footnotesize{Density fit parameters for the double
    edge model defined in eqns. 1 - \protect\ref{eq:densitymod} for the 
   density profiles shown in Fig. \protect\ref{fig:edgefits}. Column
   definitions are: (1) angular sector, (2) inner edge location, (3)
   inner density jump, (4) outer edge location (5) outer density jump 
   (6) inner, (7) middle, (8) outer slopes of the density power law. 
   The angular sectors
   are centered at the nucleus of NGC\,5846 and subtend  the angle
    from  $123^\circ$ to $180^\circ$ (NE) and from  $268^\circ$ to
    $338^\circ$ (SW), measured counterclockwise from west. 
   Uncertainties are $90\%$ confidence levels.}
}
\end{deluxetable}

%\clearpage
\begin{deluxetable}{ccccc}
\tablewidth{0pc}
\tablecaption{Temperatures Across the NE Edges\label{tab:needgespec}}
\tablehead{\colhead{Region} & \colhead{b$_{\rm i}$,a$_{\rm i}$}& 
  \colhead{b$_{\rm o}$,a$_{\rm o}$} & \colhead{$kT$} & \colhead{$\chi^2/{\rm dof}$} \\
    &(arcmin, arcmin) & (arcmin, arcmin) & (keV) & }
\startdata
NEI-2 &$0.62$,$0.72$ &$0.74$,$0.86$ &$0.64 \pm 0.05$ & $15.8/20$ \\
NEI-1 &$0.74$,$0.86$ &$0.89$,$1.03$ &$0.61 \pm 0.04$ &$23.1/22$ \\
NEI0 &$0.89$,$1.03$& $1.07$,$1.24$ & $0.62 \pm 0.03$ & $27.4/23$\\
NEI1 &$1.07$,$1.24$ &$1.29$,$1.48$ &$0.64 \pm 0.03$ & $39.5/30$\\
NEO-2 & $1.03$, $1.19$& $1.66$, $1.91$  & $0.63 \pm 0.02$ & $48.6/58$ \\ 
NEO-1 &$1.66$, $1.91$ &$2.65$, $3.05$  & $0.71 \pm 0.02$ & $69.5/66$ \\
NEO0  & $2.65$, $3.05$  & $4.24$, $4.88$ & $1.00 \pm 0.04$ &$75.4/57$ \\ 
\enddata
\tablecomments{\footnotesize{Spectral model fits in concentric
    elliptical annular regions concentric to the bounding ellipse 
    described in Fig. \protect\ref{fig:edgefits} and restricted to the 
    angular sector from $123^\circ$ to $180^\circ$ measured counter
    clockwise from west. The northeast inner edge is at the
    boundary between NEI-1 and NEI0. The northeast outer edge
    is at the boundary between NEO-1 and NEO0. The spectral 
    model is an absorbed VAPEC model with Galactic absorption and 
    abundances fixed as in Table \protect\ref{tab:tprof}}
}
\end{deluxetable}

\begin{deluxetable}{cccc}
\tablewidth{0pc}
\tablecaption{Temperatures Across the SW Edges\label{tab:swedgespec}}
\tablehead{\colhead{Region} & \colhead{r$_{\rm i}$,r$_{\rm o}$}& 
\colhead{$kT$} & \colhead{$\chi^2/{\rm dof}$} \\
    &(arcmin, arcmin) & (keV) & }
\startdata
SWI-2 &$0.95$,$1.24$ &$0.76^{+0.03}_{-0.04}$ & $44.5/34$ \\
SWI-1 &$1.24$,$1.61$ &$0.74^{+0.05}_{-0.05}$ & $11.0/19$ \\
SWI0 &$1.61$,$2.10$  & $0.94^{+0.05}_{-0.05}$ & $26.1/33$\\
SWO-1 &$2.11$, $2.74$  & $1.03^{+0.04}_{-0.04}$& $17.6/36$ \\
SWO0  & $2.74$, $3.57$ & $1.17^{+0.10}_{-0.19}$ &$25.6/27$ \\ 
SWO1  & $3.57$,$4.64$ & $1.014^{+0.07}_{-0.08}$ &$35.6/31$ \\
\enddata
\tablecomments{\footnotesize{Spectral model fits in concentric
    circular annular regions centered on NGC5846's nucleus and 
    restricted to lie in the angular sector from 
    $268.2^\circ$ to $338.0^\circ$ measured counter
    clockwise from west. The southwest inner edge is at the
    boundary between SWI-1 and SWI0. The southwest outer edge
    is at the boundary between SWO-1 and SWO0. The spectral 
    model is an absorbed VAPEC model with Galactic absorption and 
    abundances fixed as in Table \protect\ref{tab:tprof}}
}
\end{deluxetable}

\begin{deluxetable}{ccccccc}
\tablewidth{0pc}
\tablecaption{Cold Front Analyses\label{tab:edgeratios}}
\tablehead{\colhead{Edge} & \colhead{$r_{\rm edge}$}& 
\colhead{$n_i/n_o$} & \colhead{$T_i/T_o$} & $p_i/p_o$ & Mach & v \\
    & (kpc) &  & & & & (km\,s$^{-1}$) }
\startdata
NE$_{\rm outer}$ &$19.8$ &$2.9^{+0.5}_{-0.4}$ &$0.71^{+0.5}_{-0.5}$
 &$2.1^{+0.5}_{-0.4}$ &$1.0^{+0.2}_{-0.2}$ & $520^{+90}_{-100}$ \\
SW$_{\rm outer}$ &$19.1$ &$2.1^{+0.7}_{-0.4}$ & 
 $0.88^{+0.21}_{-0.10}$ &$1.8^{+1.2}_{-0.6}$ &$0.9^{+0.4}_{-0.3}$ 
 & $510^{+220}_{-240}$\\
NE$_{\rm inner}$ &$6.7$ &$1.4^{+0.2}_{-0.2}$
 &$0.98^{+0.12}_{-0.10}$ &$1.3^{+0.4}_{-0.3}$
 &$0.6^{+0.3}_{-0.5}$ & $250^{+100}_{-190}$ \\
SW$_{\rm inner}$ &$11.0$ &$1.6^{+0.2}_{-0.2}$
 &$0.79^{+0.10}_{-0.07}$ & $1.3^{+0.4}_{-0.3}$
 &$0.6^{+0.2}_{-0.4}$ & $260^{+160}_{-190}$\\
\enddata
\tablecomments{\footnotesize{Cold front ratio analyses following 
Vikhlinin \etal (2001). The edge positions are measured from the
center of NGC\,5846. Mach numbers are relative to the speed of sound 
outside each edge.  
}}
\end{deluxetable}

\begin{deluxetable}{cccccccc}
\tablewidth{0pc}
\tablecaption{Details of the VLA Radio Observations\label{tab:radiodat}}
\tablehead{\colhead{Project} & \colhead{Observation}& \colhead{Array}&
\colhead{Frequency} & \colhead{Bandwidth}& \colhead{Integration  $^1$} 
 &\colhead{ FWHM, PA $^2$}  & \colhead{rms}  \\
   &  date &  & (GHz) &  (MHz) &  time (min)   &
 ($^{\prime \prime} \times^{\prime \prime}$, $^{\circ}$) &
           ($\mu$Jy b$^{-1}$) }
\startdata
AW202 &  Jan 1988 & B & $1.5$ & $50$ & $12$ & $5.9 \times 4.6$, $-21$ & $35$ \\
AF389 & Mar 2002 & A & $1.5$ & $50$ & $153.7$ & $1.5 \times 1.5$, $0$ & $11.2$ \\
AF142  & Feb 1987 & CnD &$4.9$& $50$ & $3$ & $8.7 \times 4.7$, $-74$ & $25$ \\ 
\enddata
\tablecomments{\footnotesize{$^1$ Integration time on source; $^2$ 
full-width half-maximum (FWHM) and position angle (PA) of the full
array.
}}
\end{deluxetable} 

\begin{deluxetable}{ccccccc}
\tablewidth{0pc}
\tablecaption{Bubble Properties\label{tab:bubbles}}
\tablehead{\colhead{Label} & \colhead{$d$}& 
\colhead{r} &\colhead{ $4pV$} &\colhead{$t_s$} &\colhead{$t_b$} &\colhead{$L_{\rm mech}$} \\
    & (kpc) & (kpc)  & ( $10^{55}$\,ergs) & (Myr) &(Myr) &($10^{41}$\ergs) }
\startdata
 ghost &$5.23$  &$1.68$ &$6$  &$11.2$ &$12.1$ & $1.5$  \\
 north &$0.75$  &$0.58$ &$1$  &$1.8$  &$1.1$  & $3.8$ \\
 south &$0.93$  &$0.58$ &$1$  &$2.2$  &$1.6$  & $2.5$ \\
\enddata
\tablecomments{\footnotesize{Columns (1) bubble label, (2) distance $d$
    from the nucleus of NGC\,5846,  (3) bubble radius,   
(4) the work needed to evacuate the
    cavity for a relativistic plasma ($\gamma = 4/3$), 
(5) bubble age for bubble rising at the speed of sound in the ambient
    gas
(6) bubble age for bubble rising buoyantly at its terminal velocity, (7) 
   the instantaneous mechanical
    power of the outburst estimated using the buoyancy timescale $t_b$
}}
\end{deluxetable}

\begin{deluxetable}{ccccc}
\tablewidth{0pc}
\tablecaption{Central Spectral Extraction Regions\label{tab:centralregs}}
\tablehead{\colhead{Label} & \colhead{RA,Dec} &  \colhead{$r$} 
& \colhead{$a_{\rm out}$, $b_{\rm out}$} & 
\colhead{$a_{\rm in}$, $b_{\rm in}$}\\
  & (J2000.0) & (arcsec) & (arcsec,arcsec) & (arcsec,arcsec)  }
\startdata
Knot$1$   &  $15:06:29.080,+01:36:27.09$  &  $1.09 $ &\ldots  & \ldots  \\
Knot$2$   &  $15:06:29.182,+01:36:29.12$  &  $1.20 $ &\ldots & \ldots  \\
Knot$3$   &  $15:06:29.007,+01:36:21.06$  &  $1.06 $ &\ldots & \ldots \\
Knot$4$   &  $15:06:28.953,+01:36:24.39$  &  $1.02 $ &\ldots & \ldots \\
Knot$5$   &  $15:06:28.850,+01:36:18.90$  &  $1.07 $ &\ldots & \ldots \\
Knot$6$   &  $15:06:29.505,+01:36:18.86$  &  $1.40 $ &\ldots & \ldots \\
Knot$7$   &  $15:06:29.473,+01:36:30.65$  &  $1.20 $ &\ldots & \ldots \\
Knot$8$   &  $15:06:29.329,+01:36:29.79$  &  $1.20 $ &\ldots & \ldots \\
Knot$9$   &  $15:06:29.078,+01:36:19.29$  &  $1.06 $ &\ldots & \ldots \\
North Rim &$15:06:29.518,+01:36:24.75$  & \ldots &$10.3$, $9.6$ &
$4.5$, $4.1$ \\
South Rim & $15:06:28.943,+01:36:13.93$ & \dots & $8.8$, $8.3$ &
 $3.7$, $3.5$ \\
Background & $15:06:29.223,+01:36:21.06$ & \ldots &$33.2$,$31.5$ & 
$19.5$, $18.4$ 
\enddata
\tablecomments{\footnotesize{Spectral regions for features in
    the central region of NGC\,5846. Column (1) region
    label, (2) spectral extraction region center, (3) radius of
    circular regions used for knots and the nucleus, 
   (4) outer and (5) inner semi-major, semi-minor axes
    of elliptical annulus regions used for the north and south rims
    and common local background. Knots $1$,$2$, $4$, $6--8$  are located only 
    in the North Rim, knot $5$ is only in the South Rim, and  
   knots $3$ and $9$ are 
   in the projected overlapping region common to both rims.   
}}
\end{deluxetable}
\clearpage
\begin{deluxetable}{cccccc}
\tablewidth{0pc}
\tablecaption{Central Features Source Counts\label{tab:netcounts}}
\tablehead{\colhead{Label} & \colhead{$0.6-0.8$\,keV} 
& \colhead{$0.8-1.1$\,keV} & \colhead{$1.1-2$\,keV}  &
  \colhead{$2-4$\,keV} \\
  & (net counts) & (net counts) &(net counts) & (net counts)}
\startdata

 Knot$1$   &$25 \pm 5$  & $64 \pm 9$ &$25 \pm 5$ & $3 \pm
  2$  \\
 Knot$2$   &$24 \pm 5$ & $53 \pm 8$  &$44 \pm 7$ & $4 \pm
 2$  \\
 Knot$3$   &$18 \pm 5$ & $44 \pm 7$ &$28 \pm 6$ & $4 \pm
 2$  \\
 Knot$4$   &$26 \pm 6$ & $43 \pm 7$  &$43 \pm 7$ & $4 \pm
 2$  \\
 Knot$5$   &$22 \pm 5$ & $36 \pm 7$  &$22 \pm 5$ & $2 \pm
 1$  \\
 Knot$6$   &$39 \pm 7$ & $85 \pm 10$ &$52 \pm 8$ & $4 \pm
 2$  \\
 Knot$7$   &$21 \pm 5$ & $43 \pm 7$ &$33 \pm 6$ & $-0.37
\pm 0.03$  \\
 Knot$8$   &$18 \pm 5$ & $58 \pm 8$ &$25 \pm 5$ & $1 \pm
 1$  \\
 Knot$9$   &$29 \pm 6$ & $56 \pm 8$ &$28 \pm 6$ & $14 \pm
 4$  \\
 North Rim &$607 \pm 29$   &$1430 \pm 44$ &$884 \pm 33$ 
  & $113 \pm 12$ \\
 South Rim & $394 \pm 24$ &$922 \pm 37$ &$505 \pm 26$ &
  $89 \pm 10$\\
\enddata
\tablecomments{\footnotesize{
Net source counts in various energy bands for the features identified in Table
\protect\ref{tab:centralregs} summed over both observations, and using the
  common local background also given in Table \protect\ref{tab:centralregs}. }
 Knots $1$,$2$, $4$, $6--8$  are located only  in the North Rim, knot $5$ is 
only in the South Rim, and  knots $3$ and $9$ are
   in the projected overlapping region common to both rims.
}
\end{deluxetable}
\begin{deluxetable}{cccc}
\tablewidth{0pc}
\tablecaption{Central Features Hardness Ratios\label{tab:hardratio}}
\tablehead{\colhead{Label} & Total Counts &
  \colhead{$({\rm FeS} - {\rm FeSS})/({\rm FeS} + {\rm  FeSS})$} &
  \colhead{$({\rm M} -{\rm  S})/({\rm S}+{\rm M} +{\rm H}) $}}
\startdata
Knot$1$ &$117$ &$0.44$ & $-0.56$\\
Knot$2$ &$125$ &$0.37$ & $-0.27$\\
Knot$3$ &$94$  &$0.43$ & $-0.36$\\
Knot$4$ &$116$ &$0.24$ & $-0.23$\\
Knot$5$ &$82$  &$0.23$ & $-0.44$ \\
Knot$6$ &$181$ &$0.37$ & $-0.40$ \\
Knot$7$ &$97$  &$0.35$ & $-0.32$\\
Knot$8$ &$102$ &$0.52$ & $-0.50$\\
Knot$9$ &$127$ &$0.32$ & $-0.44$ \\
North Rim&$3034$&$0.40$& $-0.38$\\
South Rim&$1910$&$0.40$ &$-0.42$ \\
\enddata
\tablecomments{\footnotesize{
Hardness ratios in the Fe peak and medium energy band 
for the features identified in Table \protect\ref{tab:centralregs}
using the net source counts summed over both observations, given in Table
\protect\ref{tab:netcounts}. FeSS, FeS, M, H and S denote the
$0.6-0.8$\,keV, $0.8-1.1$\,keV, $1.1 - 2.0$\,keV, $2.0 - 4.0$\,keV and
$0.6-1.1$\,keV energy bands, respectively, and total counts are the
  net source counts in the $0.6-4$\,keV energy band.}
}
\end{deluxetable}

\begin{deluxetable}{cccc}
\tablewidth{0pc}
\tablecaption{Central Spectral Models\label{tab:centralspec}}
\tablehead{\colhead{Label} & \colhead{$kT$} & 
\colhead{ $\chi^2/{\rm dof}$} & \colhead{ $L_{\rm X}$} \\
  & (keV) & & ($10^{40}$\ergs)  }
\startdata
North Rim  & $0.72^{+0.03}_{-0.03}$ & $111/65$ & $1.0$ \\
South Rim  & $0.64^{+0.02}_{-0.03}$ & $26.9/39$ & $0.65$ \\ 
Stacked Knots  & $0.72^{+0.04}_{-0.07}$ & $67.4/45$ & $0.31$ \\
Knot$6$  & $0.72^f$ & $8.0/15$& $0.057$ \\
\enddata
\tablecomments{\footnotesize{ Spectral models fit over the
    $0.5-2$\,keV energy range using an absorbed VAPEC
    model with Galactic absorption and abundances fixed as in 
    \protect\S\ref{sec:abundances}. $L_{\rm X}$ is the $0.5-2$\,keV 
   intrinsic X-ray luminosity, assuming a luminosity distance of $24.2$\,Mpc.
   Spectra for all regions other than the south rim are extracted from
   OBSID 7923 data only. The south rim is a simultaneous fit to OBSID
   7923 and 788. 
}}
\end{deluxetable}

\vfill
\eject
\end{document}